\documentclass[aip,amsmath,amssymb,nofootinbib,reprint]{revtex4-1}

\usepackage{mathtools} % Supersedes amsmath
\usepackage{amsthm}
\usepackage{graphicx}% Include figure files
\usepackage[caption=false]{subfig}
\usepackage{dcolumn}% Align table columns on decimal point
\usepackage{bm}% bold math
\usepackage{bbm}% 
\usepackage[utf8]{inputenc}
\usepackage{mathptmx}
\usepackage{physics} % Various common shortcuts
\usepackage{enumerate} % Simple ordered lists

\usepackage{tikz}

\usetikzlibrary{calc,decorations.pathreplacing,calligraphy,matrix}
\tikzset{Dotted/.style={% https://tex.stackexchange.com/a/52856/194703
    line width=\pgfkeysvalueof{/tikz/Young/dot size},
    dash pattern=on 0.001\pgflinewidth off #1,line cap=round,
    shorten <=#1},Dotted/.default=3pt,
    vdots/.style={draw=none,path picture={
     \draw let \p1=(path picture bounding box.north),
        \p2=(path picture bounding box.south) in
        [Dotted={(\y1-\y2)/4}]
        (\p1) -- (\p2);
    }},
    cdots/.style={draw=none,path picture={
     \draw let \p1=(path picture bounding box.east),
        \p2=(path picture bounding box.west) in
        [Dotted={(\x1-\x2)/4}]
        (\p2) -- (\p1);
    }},
    none/.style={draw=none},
    Young tableau/.style={matrix of math nodes,nodes in empty cells,
    nodes={draw,
        minimum size=\pgfkeysvalueof{/tikz/Young/cell size},
        inner sep=0.5pt,
        text depth=0.14cm,
        text height=0.3cm},
    column sep=-\pgflinewidth,row sep=-\pgflinewidth},
    Young/.cd,cell size/.initial=1.75em,
    dot size/.initial=1.2pt
    }

\usepackage[frozencache,cachedir={_minted-main}]{minted}
\usepackage[T1]{fontenc}
\usepackage{lmodern}

\newtheorem{theorem}{Theorem}[section]

\newcommand{\SU}{\operatorname{SU}} % Special Unitary group
\newcommand{\GL}{\operatorname{GL}} % General Linear group

\newcommand{\su}{\mathinner{\mathfrak{su}}}
\newcommand{\spl}{\mathinner{\mathfrak{sl}}}

\DeclareMathOperator{\Ad}{Ad}

\newcommand{\Gprod}[1][k]{\operatorname{G}^{\otimes #1}_{N}} % Tensor product of k copies of adjoint.
\newcommand{\irrep}{irreducible representation} % save typing, change later
\newcommand{\irreps}{irreducible representations} % save typing, change later

\newcolumntype{e}{>{\(}c<{\)}} % Automatic math mode center column

\newcommand{\kmin}{k_{\mathrm{min}}}

\makeatletter
\newcommand*\rel@kern[1]{\kern#1\dimexpr\macc@kerna}
\newcommand*\widebar[1]{%
  \begingroup
  \def\mathaccent##1##2{%
    \rel@kern{0.8}%
    \overline{\rel@kern{-0.8}\macc@nucleus\rel@kern{0.2}}%
    \rel@kern{-0.2}%
  }%
  \macc@depth\@ne
  \let\math@bgroup\@empty \let\math@egroup\macc@set@skewchar
  \mathsurround\z@ \frozen@everymath{\mathgroup\macc@group\relax}%
  \macc@set@skewchar\relax
  \let\mathaccentV\macc@nested@a
  \macc@nested@a\relax111{#1}%
  \endgroup
}
\makeatother

\begin{document}

\preprint{AIP/123-QED}

\title[Lowest Dimensional Portals to SU(\textit{N}) Exotics]{Lowest Dimensional Portals to SU(\textit{N}) Exotics}
% Force line breaks with \\

\author{Tesla Jeltema}
 \email{tesla@ucsc.edu}
\affiliation{ 
Department of Physics, University of California and Santa Cruz Institute for Particle Physics, University of California, 1156 High St, Santa Cruz, CA 95064, USA}
\author{Stefano Profumo}
 \email{profumo@ucsc.edu}
\affiliation{ 
Department of Physics, University of California and Santa Cruz Institute for Particle Physics, University of California, 1156 High St, Santa Cruz, CA 95064, USA}
\author{Jaryd F. Ulbricht}
 \email{julbrich@ucsc.edu}
\affiliation{ 
Department of Physics, University of California and Santa Cruz Institute for Particle Physics, University of California, 1156 High St, Santa Cruz, CA 95064, USA}

\date{\today}% It is always \today, today,

\begin{abstract}
New matter fields charged under the strong nuclear force would have dramatic phenomenological implications. Here, we systematically explore how these new states, which we postulate belong to some representation of \(\SU(3)\) of quantum chromo-dynamics, could interact with Standard Model fields: We analyze  all lowest-dimensional ``portal'' operators for any \(\SU(3)\) representation and, motivated by grand unification, we extend our results to \(\SU(N)\), for $N>3$. We provide a publicly available {\tt python} code, ``{\tt tessellation}'' that automatically constructs said lowest-dimensional portal operators for any new exotic matter field charged under \(\SU(N)\) for any $N$.
\end{abstract}

\maketitle

\section{Introduction}
\label{sec:intro}
% what is confinement
In the Standard Model of particle physics, matter fields are spin-1/2 Dirac fermions. Strong nuclear interactions are highly successfully modeled by an unbroken local gauge symmetry with gauge group \(\SU(3)\). Matter fields that are strongly interacting belong to the fundamental (\(\bm{3}\)) representation of \(\SU(3)\), while their antimatter counterparts to the antifundamental (\(\bm{\bar{3}}\)) representation of \(\SU(3)\). In terms of Dynkin labels, strongly interacting matter fields belong to the (1,0) representation, and antimatter, strongly interacting fields to the (0,1) representation of \(\SU(3)\). The spin-1, massless force carriers, ``gluons'', belong to the adjoint (\(\bm{8}\)) representation, the (1,1).

Grand unification schemes, where two or more of the Standard Model gauge groups arise from symmetry breaking, at some very high energy scale, of larger ``unified'' gauge groups, have a similar structure. For instance, in Pati-Salam models based on \(\SU(4) \times \SU(2)_{\mathrm{L}} \times \SU(2)_{\mathrm{R}}\) or \(\SU(4) \times \SU(2)_{\mathrm{L}} \times \SU(2)_{\mathrm{R}} / \mathbb{Z}_{2}\) \cite{PhysRevD.10.275}, matter fermions charged under strong interactions are accommodated in the fundamental representations \(\bm{4} \cong \qty(1,0,0)\) and \(\bm{\bar{4}} \cong \qty(0,0,1)\), with additional exotic states in the \(\bm{6} \cong \qty(0,1,0)\) representation; in Georgi-Glashow \(\SU(5)\) grand unified theories \cite{Georgi:1999wka} matter fermions are accommodated in the \(\bm{5} \cong \qty(1,0,0,0)\), \(\bm{10} \cong \qty(0,1,0,0)\), and their conjugate representations \(\bm{\bar{5}}\) and \(\bm{\bar{10}}\); in \(\SU(6)\), Standard Model matter fermions are accommodated in the \(\bm{6} \cong \qty(1,0,0,0,0)\) and \(\bm{15} \cong \qty(0,1,0,0,0)\) and their conjugate representations \(\bm{\bar{6}}\) and \(\bm{\bar{15}}\), with possible additional beyond-the-Standard-Model matter fermions in the \(\bm{20} \cong \qty(0,0,1,0,0)\) (see e.g. \cite{PhysRevD.71.095013} and \cite{FUKUGITA1982369}).

In constructing beyond-the-Standard-Model theories, new exotic states are assigned charges, or representations, under the Standard Model gauge groups (or under grand unified gauge groups as the case may be). If these exotic states belong to non-trivial gauge group representations it is of paramount importance for phenomenology to establish which interaction terms are allowed by gauge invariance to exist in the model's Lagrangian. Such interaction terms, of course, may not be renormalizable. If they are not, it is implicitly assumed that some high-scale physics is effectively integrated out, yielding the higher dimensional operators. In either case, the most important operators at low energies are those with the lowest mass dimension. In this study, we are precisely interested in which arrangement of matter fields and force mediators (in the adjoint representation) are needed to produce the {\em lowest mass dimension gauge-invariant operator} (``portal'') between any given new exotic state and Standard Model fields.

In a non-perturbative regime, the question we address can be recast as the question of which arrangements of ``quarks'', ``antiquarks'' and ``gluons'' are needed to dress a given exotic state into a gauge-singlet, color-less combination: thus the question can be recast, borrowing the notion of ``valence gluon'' \cite{Chanowitz:1983ci, Buccella:1985cs}, as what is the smallest SU($N$) color-less bound state (see the recent study \cite{Profumo:2020zgi}).

From a group theoretic standpoint, we can reformulate the question as: given a gauge group SU($N$) and a new exotic state, \(X\), belonging to some representation of SU($N$), what is the most economical (in the sense of mass dimension of the corresponding operator) tensor product of fundamental and adjoint representations of SU($N$) that contains the trivial representation? Since matter fermions have higher mass dimension than spin-1 force carriers in the adjoint representation, one seeks to minimize the number of matter fermions. Since it is necessarily the case that a tensor product containing the trivial representation has vanishing $N$-ality, it is then obvious that the solution includes a single tensor product with the one fundamental representation $Q_i$ that takes the product $X\otimes Q_i$ to $N$-ality equal to zero, with as many copies $k$ of the adjoint representation $G_N$ as needed for the product $X\otimes Q_i \otimes G_N^{\otimes k}$ to eventually contain the trivial representation.

Technically, the intricacy of this problem lies in decomposing the {\em k-fold tensor product of adjoint representations} into a direct sum of \irreps{} of \(\SU(N)\). While some general results are known, for example in the context of the Littlewood-Richardson rule (see e.g. \cite{GASHAROV1998451}), we are not aware of any general result on the product of several copies of the adjoint representation. Here, we provide first, in the following section \ref{sec:tesla}, a visually simple, algorithmic approach for building the gauge invariant operator, given the $N$-ality zero combination $X\otimes Q_i$, and provide examples and complete formulae for SU($N$) with $N=3,\ 4,\ 5$. We then discuss in sec.~\ref{sec:adjoint} the problem of what is the minimal $k$ such that $\mathbbm{1} \subset X \otimes G^{\otimes k_{\mathrm{min}}}_{N}$, or, equivalently, $\bar{X} \subset G^{\otimes k_{\mathrm{min}}}_{N}$. The full problem of finding the minimal $k$ such that $\mathbbm{1} \subset X\otimes Q_i \otimes G_N^{\otimes k}$ is then solved in sec.~\ref{sec:matter}. Because, especially for large $N$ and large-dimensional representations for the exotic field $X$, the calculation is rather complicated, we describe in sec.~\ref{sec:code} a python code, {\tt tessellation}, which we make publicly available, to perform the needed calculations. We collect in two appendices the needed mathematical details, and a short user's manual for the {\tt tessellation} code.

\section{An algorithmic approach for building the gauge invariant operator}\label{sec:tesla}

In this section, we present an algorithmic approach to finding the minimum number of copies \(\kmin\) of the adjoint representation $G_N$ needed for the product with a representation with $N$-ality zero to contain the trivial representation; we believe this discussion will provide an intuitive understanding of the problem. We explicitly solve through $N=4$, outline considerations for extending to $N>4$, and present formulae for $N=3, 4, 5$ in Table~\ref{tab:examples}. We also outline some considerations for arbitrary representations with non-zero $N$-ality before proceeding to a full solution in the following sec.~\ref{sec:adjoint}.

\begin{figure}[h]
\begin{center}
\includegraphics[width=\columnwidth]{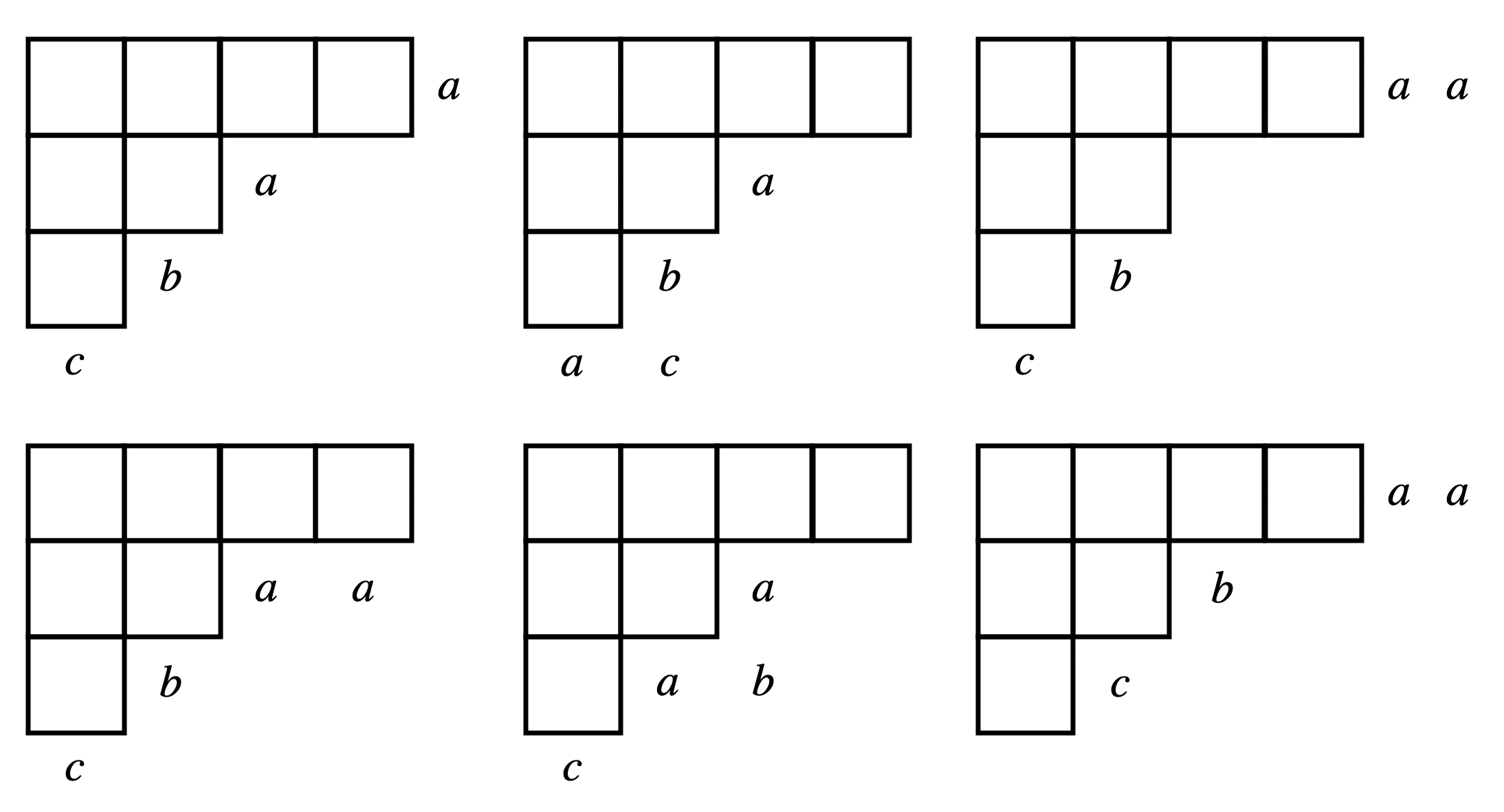}
\end{center}
\caption{Representations resulting from the tensor product of the representation (2,1,1) of SU(4) with the adjoint representation of SU(4), (1,0,1), whose Young tableaux boxes are labeled by the letters $a,\ b$ and $c$. \label{fig:adding}}
\end{figure}

We start by considering a representation with Dynkin label $\qty(p_1,...,p_{N-1})$. We then take the tensor product of this representation with the adjoint representation, whose Young tableau contains $N$ boxes, and build new tableaux (see e.g. Ref.~\cite{Slansky:1981yr} for a general discussion of Young tableaux in the context of tensor products of group representations), illustrated for instance, for $N=4$, in Fig.~\ref{fig:adding}.  Generically, the allowed configurations include placing one box on each row or placing two boxes on the same row and one box on all but one of the other rows.  The former possibility simply lengthens every row by one, effectively mapping $(p_1,...,p_{N-1}) \rightarrow (p_1,...,p_{N-1})$ to itself upon multiplication by the adjoint representation, and thus adds a copy of the adjoint representation without getting closer to a representation containing the trivial representation. We will, therefore, only consider steps which add two boxes to one of the rows.  Among these possibilities it is also never useful to add two boxes to the top row; in fact, the best step we can take is to add two boxes to the lowest row possible while maintaining rows which reduce in length from top to bottom. This allows us to complete at least one column and minimizes the steps needed to complete additional columns (see also the following sec.~\ref{sec:adjoint} for a formal proof of this statement).

Starting from an initial representation with $N$-ality equal to zero, or which we have brought to $N$-ality of zero by tensor product with the appropriate fundamental representation (see sec.~\ref{sec:matter}), we ask what is the minimum number copies of the adjoint representation, $\kmin$, to get to a direct sum of \irreps{} that includes the trivial representation, effectively, and visually, filling in a full rectangle of height $N$ in our Young tableau.

\begin{figure}
    \subfloat[\label{fig:p412a}]{%
    \includegraphics[width=0.45\linewidth]{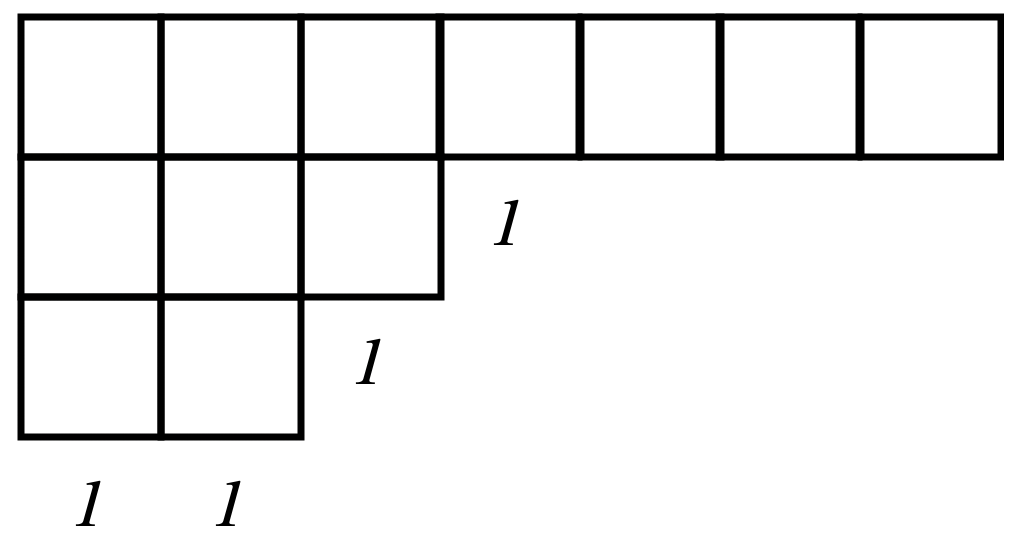}%
    }\hfill
    \subfloat[\label{fig:p412b}]{%
    \includegraphics[width=0.45\linewidth]{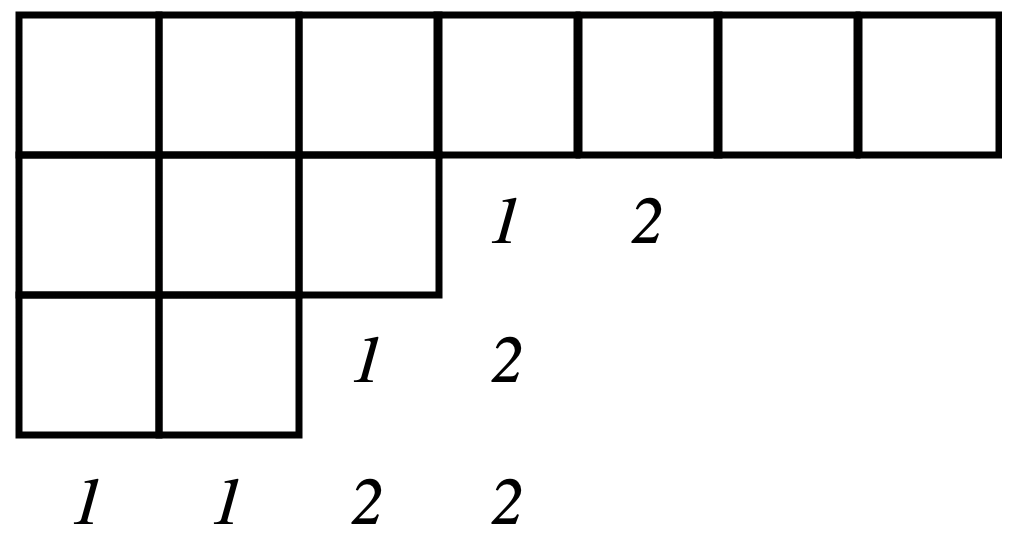}%
    }
    
    \subfloat[\label{fig:p412c}]{%
    \includegraphics[width=0.45\linewidth]{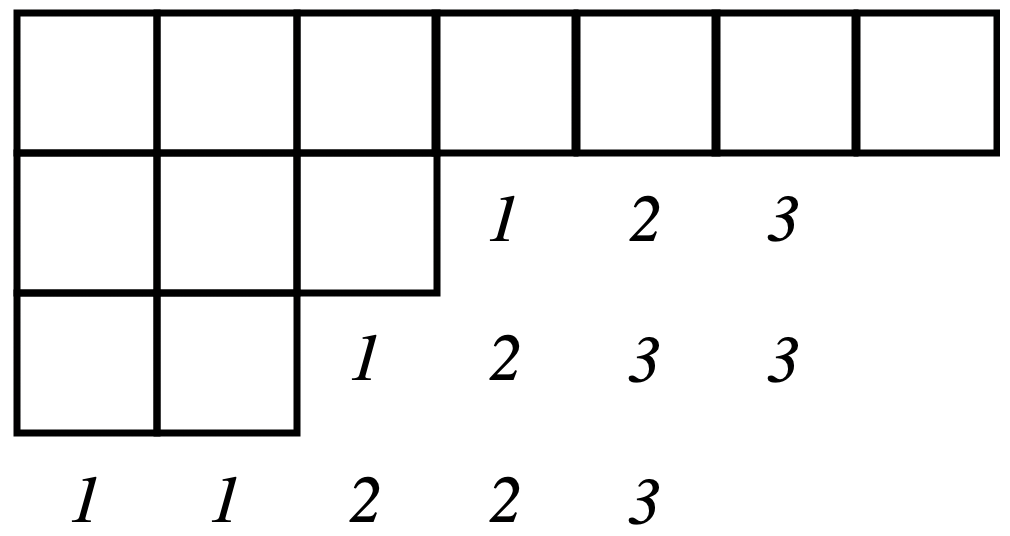}%
    }\hfill
    \subfloat[\label{fig:p412d}]{%
    \includegraphics[width=0.45\linewidth]{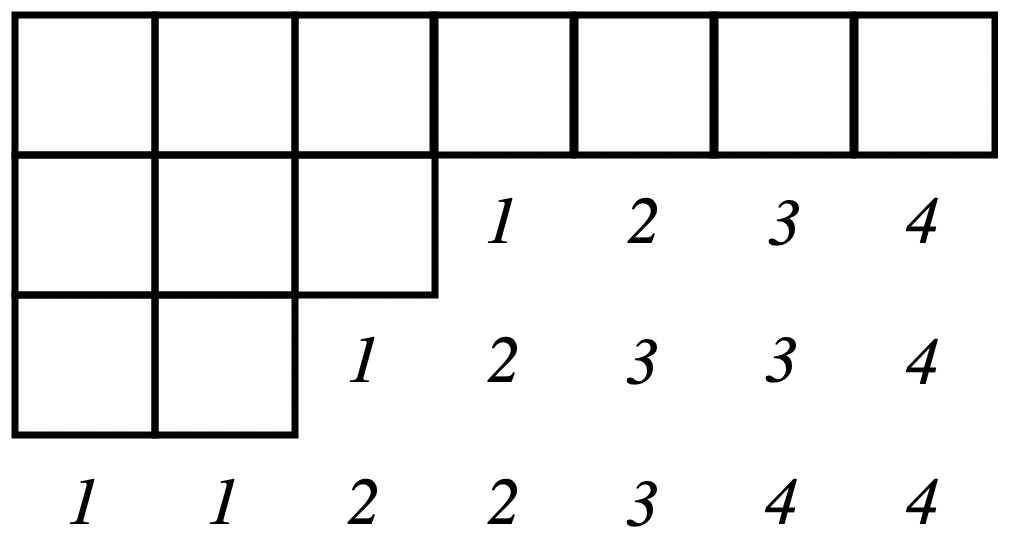}%
    }
    \caption{An example for $N=4$ of a representation, here (4,1,2), which can be completed in the minimum possible number of steps.\label{fig:p412}}
\end{figure}

We take advantage of the fact that the adjoint representation is self dual. It is convenient to choose as the starting point whichever of $(p_1,p_2,...,p_{N-1})$ or $(p_{N-1},p_{N-2},...,p_1)$ ensures $p_1 \ge p_{N-1}$.  In step 1, we then add two boxes on the bottom row and one box on each row except the top row.  The ordering with $p_1 \geq p_{N-1}$ ensures that this is an allowed step unless $p_1 = p_{N-1} = 0$.  What happens in the case of $p_1 = 0$ is described later.  This step has the effect of reducing both $p_1$ and $p_{N-1}$ by one while leaving the other $p_i$ unchanged, $(p_1,p_2,...,p_{N-1}) \rightarrow (p_1-1,p_2,...,p_{N-1}-1)$ (see Fig.~\ref{fig:p412a}).  We can then repeat this process $p_{N-1}$ times until $p_{N-1}=0$ (see Fig.~\ref{fig:p412b}).  Thus the first stage in our algorithm is:\\

\indent (1) Take $p_{N-1}$ steps (i.e. tensor products with the adjoint representation), adding two boxes to the bottom row and one box to every row except the top, reducing $(p_1,p_2,...,p_{N-1})$ to $(p_1-p_{N-1},p_2,...,0)$.\\

With $p_{N-1}=0$, the next step must add two boxes to row $N-1$ to ensure that it is longer than row $N$.  Assuming our new $p_1$ (equal to $p_1 - p_{N-1}$) is greater than zero, we will also add one box each to the other rows except the top row.  This step reduces $p_1$ by one, reduces $p_{N-2}$ by one, and increases $p_{N-1}$ to one (see Fig.~\ref{fig:p412c}). In the following step we can again add two boxes to the bottom row, reducing $p_{N-1}$ again to zero, reducing $p_1$ by one, and leaving the other $p_i$ unchanged (see Fig.~\ref{fig:p412d}).  Thus it takes two steps to reduce $p_{N-2}$ by one and have $p_{N-1}=0$. This set of two steps may be repeated, while the first term $p_1$ is greater than zero, until $p_{N-2} = p_{N-1} = 0$ after $2p_{N-2}$ additional steps.  At this point the lowest row on which we can place two boxes is $N-2$, which reduces the size of $p_{N-3}$.  Reducing $p_{N-3}$ by one and returning to $p_{N-2} = p_{N-1} = 0$ requires a set of three steps as we iteratively increase $p_{N-2}$ and $p_{N-1}$ to one and then bring them back to zero (i.e. we add two boxes first to row $p_{N-2}$, then to row $p_{N-1}$, and finally to $p_{N}$).  Thus the second stage in our algorithm is:\\

\indent (2) While $p_1 > 0$, iteratively place two boxes on the next to lowest row followed by the lowest row until  $p_{N-2} = p_{N-1} = 0$ after $2p_{N-2}$ steps; then follow a similar process of $3p_{N-3}$ steps to reduce to $p_{N-3} = p_{N-2} = p_{N-1} = 0$, etc.  The total steps taken from the starting representation is $p_{N-1} + 2p_{N-2} + 3p_{N-3} + ...$ and the value of first term is now $p_1 - p_{N-1} - 2p_{N-2} - 3p_{N-3} - ...$.

How this process concludes depends on whether the starting value of $p_1$ is equal to, greater than, or less than $$\sum_{i=2}^{N-1}(N-i)p_i.$$  We consider each of these in turn.  If $p_1 = \sum_{i=2}^{N-1}(N-i)p_i$ then we reduce $p_1$ to zero just as all of the other $p_i$ go to zero. Here we form a complete rectangle (trivial representation) in the minimum possible number of steps given the number of missing boxes in the starting state compared to a rectangle of height $N$ and length $\sum_{i=1}^{N}p_i$.

%\begin{widetext}
\begin{equation}
\begin{aligned}
    k = p_1 = \sum_{i=2}^{N-1}(N-i)p_i = \frac{\sum_{i=1}^{N-1}(N-i)p_i}{N} \\ \textrm{for} \hspace{6pt} p_1 = \sum_{i=2}^{N-1}(N-i)p_i
    \end{aligned}
\end{equation}
%\end{widetext}

\begin{figure}[h]
\begin{center}
\includegraphics[width=0.9\linewidth]{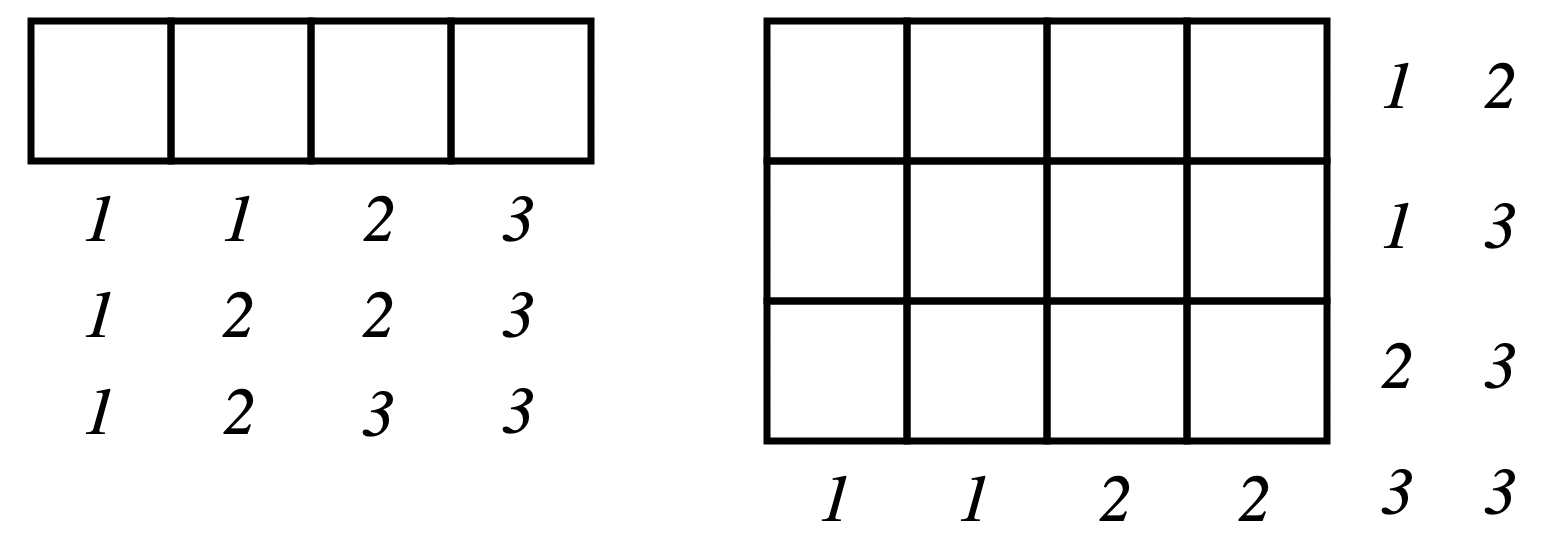}
\end{center}
\caption{Examples for $N=4$ of the symmetric  representations, (4,0,0) and (0,0,4), which can be completed in the same number of steps.\label{fig:p400}}
\end{figure}

If $p_1 > \sum_{i=2}^{N-1}(N-i)p_i$, we again complete the rectangle (i.e. get to a product of representations containing the trivial representation) in the minimum possible number of steps.  This situation implies that we can bring all of the $p_i$ with $i>1$ to zero with some of the length of $p_1$ remaining.  In this case, the remainder of $p_1$, which we shall call  $p_1'$ is necessarily a multiple of $N$ given the starting condition of $N$-ality zero, and we have a state of the form $(p_1'=jN, 0, ..., 0)$.  This representation can be reduced to the trivial representation in $(N-1)j$ steps (see e.g. the left panel of Fig.~\ref{fig:p400}).  The total number of steps from the starting representation in this case is

\begin{equation}
    \begin{aligned}
    k = \sum_{i=2}^{N-1}(N-i)p_i + \frac{(N-1)(p_1-\sum_{i=2}^{N-1}(N-i)p_i)}{N}\\ = \frac{\sum_{i=1}^{N-1}(N-i)p_i}{N} \hspace{12pt} \textrm{for} \hspace{6pt} p_1 > \sum_{i=2}^{N-1}(N-i)p_i.
    \end{aligned}
\end{equation}

\begin{figure}[h]
\begin{center}
\includegraphics[width=0.4\linewidth]{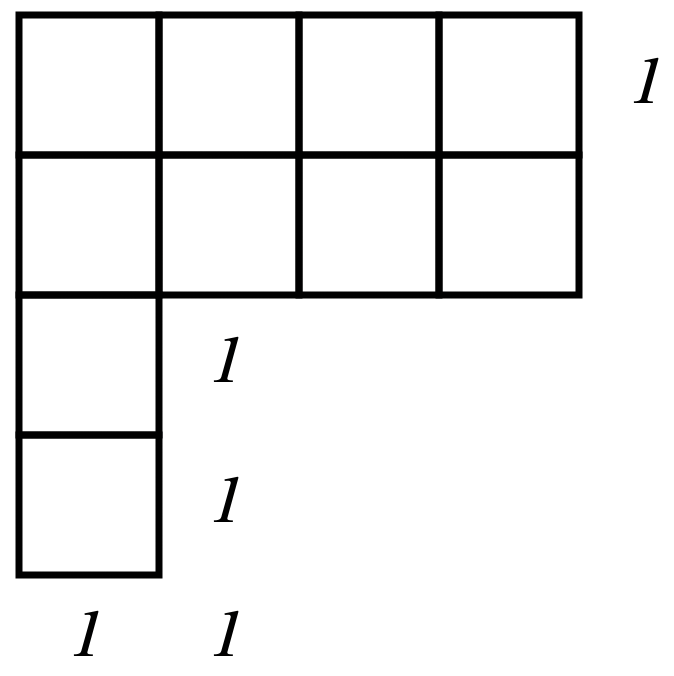}\hspace*{0.5cm}
\includegraphics[width=0.45\linewidth]{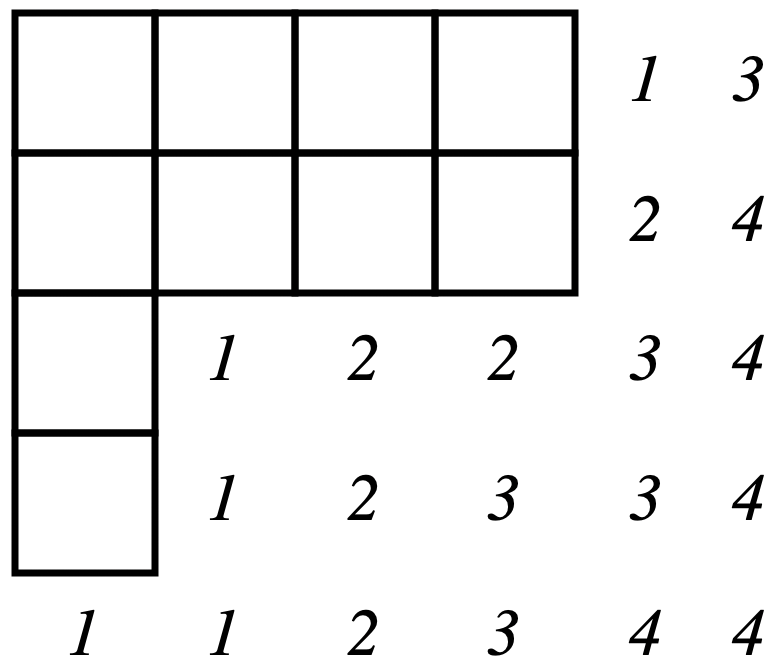}
\end{center}
\caption{Example for $N=5$ of a representation, here (0,3,0,1), for which completion requires expanding beyond the original length.} \label{fig:p0301}
\end{figure}

If $p_1 < \sum_{i=2}^{N-1}(N-i)p_i$, then the $p_1$ term reaches zero before all of the other rows are zero.  This situation requires that we expand beyond the bounds of our initial rectangle and take more than the minimum possible number of steps.  Given the reordering we did in the beginning, we know that $p_{N-1}$ is either zero or one at this point. In the step after the first term reaches zero, we add one block to row 1 increasing $p_1$ to one and reducing $p_2$ by one (see Fig.~\ref{fig:p0301}).  The rest of this step looks just like the previous ones we have taken; if $p_{N-1}=1$, this term is reduced to zero and if it is not, the first non-zero term is reduced by one.  The following step starts with $p_1=1$ and proceeds just like our previous steps reducing $p_1$ back to zero.  Thus the effect of adding steps beyond the minimum (adding columns on to our rectangle) is to reduce $p_2$ by one and add two steps above the minimum for each extra column.  We can then proceed as before reducing each of the $p_i$ in turn with the difference that when we get to reducing the $p_2$ term its value will be $p_2 - N_{extra}$, where $N_{extra}$ is the number of extra columns.  For $p_1 < \sum_{i=2}^{N-1}(N-i)p_i$, the third stage in our algorithm is:\\

\indent (3) If $p_1=0$, add a box to row 1 increasing $p_1$ to one and reducing $p_2$ by one. Add the other $N-1$ boxes on the lowest rows possible, reducing the value of the largest $i$ term $p_i$ that is not zero by one.\\

The total number of steps taken, assuming $p_2-N_{extra}$ greater than zero until the final step, is then
\begin{equation}
    \begin{aligned}
    k = (N-2)(p_2-N_{extra}) + \sum_{i=3}^{N-1}(N-i)p_i\\ = \frac{\sum_{i=1}^{N-1}(N-i)p_i}{N} + N_{extra}\\ \textrm{for} \hspace{6pt} p_1 < \sum_{i=2}^{N-1}(N-i)p_i.
    \end{aligned} \label{eqn:ncase3}
\end{equation} 
Solving for the number of extra columns gives
\begin{equation}
    N_{extra} = \frac{\sum_{i=2}^{N-1}(N-i)p_i - p_1}{N} \label{eqn:nextra}
\end{equation} 
Substituting back Eqn.~(\ref{eqn:nextra}) in to Eqn.~(\ref{eqn:ncase3}) gives
\begin{equation}
    \begin{split}
    k &= \frac{N-2}{N} p_1 + \frac{2}{N} \sum_{i=2}^{N-1}(N-i)p_i\\[2mm]
    &= \frac{2}{N} \sum_{i=1}^{N-1}(N-i)p_i - p_1\\[2mm] &p_1 < \sum_{i=2}^{N-1}(N-i)p_i, \;\; p_2 \ge \frac{1}{2}\left[\sum_{i=3}^{N-1}(N-i)p_i - p_1\right].
    \end{split} \label{eqn:n3final}
\end{equation} 

The total number of steps in this case is twice the minimum possible minus $p_1$, with $p_1$ being the number of steps we can take before adding extra columns. The condition on the size of $p_2$ comes from finding the value of $p_2$ such that it just reaches zero when the other $p_i$ all reach zero.  In this case, $k = \sum_{i=3}^{N-1}(N-i)$. At this point, we have fully solved for $N\le4$, since for $N=4$, $p_3 = p_{N-1} \le p_1$ by virtue of our starting representation and the condition on $p_2$ is always satisfied.

For $N\ge5$, there is the possibility that we arrive at a representation with zeros in the first two (or more) terms while the other $p_i$ remain non-zero, e.g. $(0,0,p_3',p_4')$.  An easy method to see the number of steps remaining in this case is to consider the ``symmetric'' case, e.g. $(p_4',p_3',0,0)$, which will require the same number of steps to reduce (see Fig.~\ref{fig:p400} for one example).  Here we know that $p_{N-1}'$ is either zero or one again by virtue of having started with $p_1 \ge p_{N-1}$.  This means our ``symmetric'' representation automatically has $p_1 < \sum_{i=2}^{N-1}(N-i)p_i$ and the solution is given by Eqn.~\ref{eqn:n3final}.  The total number of steps taken is then the sum of the number of steps to get from an initial state $(p_1,p_2,...,p_{N-1})$ to $(0,0,p_3',...,p_{N-1}')$ plus the number of steps to reduce $(p_{N-1}', ..., p_3',0,0)$. 

Explicit formulae for $k_{min}$ for $N=3, 4, 5$ are given in Table~\ref{tab:examples}.  For $N=6$, there is the possibility of arriving at representations with three leading zeros $(0,0,0, p_4',p_5')$ or of the form $(0,0,p_3',0,0)$.  The former can again be solved through symmetry.  In the latter case, because $N$-ality is zero $p_3' = jN/(N-3) = 2j$, for $N=6$ and where $j$ is a positive integer.  To reduce from here $(N-3)j = 3j$ steps are needed. For $N>6$, additional cases emerge, and we seek a fully general solution in the next section.

The following section also presents a general solution for non-zero $N$-ality.  Here we discuss a few general considerations for this case.  For a general representation $X$ with $N$-ality equal to $t$, we can bring the initial representation to $N$-ality of zero in one step by adding a fundamental representation of $N-t$ boxes.  Adding these boxes to the lowest rows possible (while maintaining rows of decreasing length) will result in the representation which will then fill most efficiently (see Fig~\ref{fig:matterfieldexamples}).  If the term $p_t \neq 0$, boxes are added to the rows below row $t$ and $p_t$ is reduced by one.  The resulting $N$-ality zero representation will be $(p_1, ..., p_t - 1, ..., p_{N-1})$. What happens if $p_t=0$ depends on if the surrounding terms are also zero, but in general one of the zero terms increases to one while the non-zero terms above and below decrease by one. For example, if $p_t=0$ and $p_{t+1}, p_{t-1} \neq 0$, then the resulting $N$-ality zero representation will be $(p_1, ..., p_{t-1}-1, 0, p_{t+1}-1, ..., p_{N-1})$. 

\begin{table*}
\setlength{\tabcolsep}{0.5em}
{\renewcommand{\arraystretch}{2.0}
\begin{tabular}{|e|e|e|e|}
\hline
N & p_1 \ge \sum_{i=2}^{N-1}(N-i)p_i & p_1 < \sum_{i=2}^{N-1}(N-i)p_i & p_1 < \sum_{i=2}^{N-1}(N-i)p_i \\
 & & p_2 \ge \frac{1}{2}\left[\sum_{i=3}^{N-1}(N-i)p_i - p_1\right] & p_2 < \frac{1}{2}\left[\sum_{i=3}^{N-1}(N-i)p_i - p_1\right] \\
\hline
3 & \frac{2p_1+p_2}{3} & \frac{p_1+2p_2}{3} & - \\
\hline
4 & \frac{3p_1+2p_2+p_3}{4} & \frac{2p_1+4p_2+2p_3}{4} & - \\
\hline
5 & \frac{4p_1+3p_2+2p_3+p_4}{5} & \frac{3p_1+6p_2+4p_3+2p_4}{5} & \frac{2p_1+4p_2+6p_3+3p_4}{5}\\
\hline
\end{tabular}
}
\caption{\label{tab:examples} Minimum number of steps, $k_{min}$, for a state with $N$-ality of zero and $p_1 \ge p_{N-1}$.  For a state with $p_{N-1} > p_1$, the number of steps is that of the symmetric state $(p_{N-1},p_{N-2},...,p_1)$.}
\end{table*}

\section{Solution for General \textit{N}}\label{sec:adjoint}

We begin by decomposing the tensor product of \(k\) copies of the adjoint representation. Because the group \(\SU(N)\) is compact and simply connected a representation on \(\SU(N)\) induces a unique representation on the algebra \(\su(N)\). By Weyl's theorem on complete reducibility of semisimple Lie algebras over a field of characteristic zero the tensor product of \(k\) copies of the adjoint representation then decomposes into a direct sum of \irreps{} on \(\SU(N)\). After having obtained this decomposition it is simple to see that for any \irrep{} \(X\),
\begin{equation} \label{eq:trivialcondition}
    \bar{X} \subset \Gprod \implies \mathbbm{1} \subset X \otimes \Gprod,
\end{equation}
where \(\bar{X}\) is the conjugate representation to \(X\). This allows us to construct an expression for the minimum number of copies of the adjoint, \(\kmin\), such that the above condition holds for any \(X\).

We then include matter fields by letting \(X \to X \otimes Q_{i}\), where \(i\) is fixed by requiring this tensor product have vanishing \(N\)-ality. This generally produces a direct sum of \irreps{}, of which we only require the single \irrep{} that gives the smallest \(\kmin\). To this end we develop an algorithmic procedure for determining the optimum \irrep{} in \(X \otimes Q_{i}\) such that
\begin{equation}
    \mathbbm{1} \subset X \otimes Q_{i} \otimes \Gprod[\kmin].
\end{equation}

\subsection{Adjoint Product Decomposition} \label{subsec:AdProd}

Consider the highest weight of \(G_{N}\), the Dynkin label of the adjoint representation:
\begin{equation}
\bar{\lambda} \equiv (1, \underbrace{0, \ldots, 0}_{N-3}, 1).
\end{equation}
The highest weight of the highest \irrep{} of \(G^{\otimes k}_{N}\), \(\Lambda\), results from adding this weight to itself \(k\) times:
\begin{equation}
    \Lambda \equiv k \bar{\lambda} = (k, \underbrace{0, \ldots, 0}_{N-3}, k).
\end{equation}
This weight belongs to one and only one \irrep{} in \(\Gprod\), which we denote \(V_{\Lambda}\). To \(V_{\Lambda}\) we associate the tableau
\begin{equation}
    V_{\Lambda} \cong \vcenter{\hbox{
\begin{tikzpicture}
\node[Young tableau](yt1){ & |[cdots]| & & & |[cdots]| & \\
 & |[cdots]| & \\
|[vdots]| & |[none]| & |[vdots]| \\
 & |[cdots]| & \\};
\draw[very thick, decoration={calligraphic brace, amplitude=6pt}, decorate] (yt1-4-1.south west) -- (yt1-1-1.north west) node[midway, above=1ex, sloped]{\(N - 1\)};
\draw[very thick, decoration={calligraphic brace, amplitude=6pt}, decorate] (yt1-1-1.north west) -- (yt1-1-3.north east) node[midway, above=1ex, sloped]{\(k\)};
\draw[very thick, decoration={calligraphic brace, amplitude=6pt}, decorate] (yt1-1-4.north west) -- (yt1-1-6.north east) node[midway, above=1ex, sloped]{\(k\)};
\end{tikzpicture}}}
\end{equation}
To generate the other \irreps{} in \(\Gprod\) we move boxes from the top rows of \(V_{\Lambda}\) to lower rows, and require that the resulting tableau remain valid (i.e. the number of boxes in each row are weakly decreasing). For example, with \(k = 3\) and \(N = 4\) we can do the following moving only one box:
\begin{equation*}
    \vcenter{\hbox{
    \begin{tikzpicture}
    \node[Young tableau](yt){ & & & & & \star \\
     & & & |[dotted, thick]| \star \\
     & & \\
     |[dotted, thick]| \star \\
     };
    \draw[thick, ->] (yt-1-6.south) to[out=270, in=0] (yt-2-4.east);
    \draw[thick, ->] (yt-1-6.south) to[out=270, in=0] (yt-4-1.east);
    \draw[thick, ->] (yt-3-3.south) to[out=270, in=0] (yt-4-1.east);
    \end{tikzpicture}}}
\end{equation*}
More generally, consider moving \(i_{jk}\) boxes from row \(j\) to row \(k\) with \(j < k\). To this procedure we associate a totally antisymmetric \(N \times N\) matrix:
\begin{equation} \label{eq:M}
    M = \mqty*(0 & i_{12} & i_{13} & \ldots & i_{1N} \\
    - i_{12} & 0 & i_{23} & \ldots & i_{2N} \\
    - i_{13} & - i_{23} & 0 & \ldots & i_{3N} \\
    \vdots & \vdots & \vdots & \ddots & \vdots \\
    - i_{1N} & - i_{2 N} & - i_{3 N} & \ldots & 0).
\end{equation}
The number of boxes \textit{removed} from row \(j\) is
\begin{equation}
    m_{j} \equiv \sum^{N}_{n = 1} M_{j n}.
\end{equation}
The Dynkin label of the resulting \irrep{} is then given by taking \(\Lambda\) and for each row \(j\) subtracting the \(m_{j}\) boxes removed from that row and adding the \(m_{j + 1}\) boxes removed from the row below it. We write this as a \textit{correction} vector \(\Delta p\), giving \(\Lambda - \Delta p\) as the Dynkin label of the resulting \irrep{}.
\begin{subequations}
    \begin{align}
    \Delta p_{jk} \coloneqq& \; m_{j} - m_{k}\\[2mm]
    \Delta p_{j} \equiv& \; \Delta p_{j \: j+1} = m_{j} - m_{j + 1}\\[2mm]
    \Delta p \equiv& \; \qty(\Delta p_{1}, \Delta p_{2}, \ldots, \Delta p_{N - 1}).
\end{align}
\end{subequations}
This defines a mapping from the space of matrices which contain all \(M\) to the \irreps{} in \(\Gprod\). The multiplicities of these \irreps{} is irrelevant, as long as they are greater than zero. In Appendix~(\ref{app:adjointprod}) we derive an analog of Schur-Weyl duality for tensor products of the adjoint representation of \(\SU(N)\) and show that this mapping is subjective. The result is:
\begin{equation}
    \Gprod \cong \bigoplus_{\lambda} V^{\oplus m_{\lambda}(G_{N})}_{\lambda},
\end{equation}
where the sum is over dominant weights \(\lambda\), \(V_{\lambda}\) are \irreps{} of \(\SU(N)\) labeled by \(\lambda\), and the multiplicity \(m_{\lambda}(G_{N})\) is
\begin{equation}
    m_{\lambda}(G_{N}) = \sum^{k}_{n = 0} \sum_{\mu_{n}} \sum_{\nu_{n}} \qty(-1)^{k - n} \binom{k}{n} m_{\mu_{n}} m_{\nu_{n}} c^{\lambda}_{\mu_{n} \bar{\nu}_{n}},
\end{equation}
where the second and third sums are over all partitions \(\mu_{n}\) and \(\nu_{n}\) of an integer \(n\), \(m_{\mu_{n}}\) and \(m_{\nu_{n}}\) are the dimensions of the \irreps{} of the symmetric group \(\mathfrak{S}_{n}\) labeled by \(\mu_{n}\) and \(\nu_{n}\) respectively, \(c^{\lambda}_{\mu \nu}\) are the Littlewood-Richardson coefficients, and \(\bar{\nu}_{n}\) is the conjugate partition to \(\nu_{n}\). These \(m_{\lambda}(G_{N})\) are necessarily non-zero positive integers for all \(\lambda = \Lambda - \Delta p\).

The \(m_{j}\) and \(\Delta p(M)\) obey the following identities (from now on let the dependence on \(M\) be implicit):
\begin{subequations}
\begin{align}
    \sum^{N}_{j = 1} m_{j} =& \; 0, \label{eq:msum}\\[2mm]
    \sum^{N-1}_{j = n} \Delta p_{j} =& \; m_{n} - m_{N}, \label{eq:deltapjsum}\\[2mm]
    \sum^{N-1}_{j = 1} j \Delta p_{j} =& \; - N m_{N}, \label{eq:jdeltapjsum}\\[2mm] 
    \sum^{n}_{j = 1} m_{j} \geq& \; 0 \quad \forall n \leq N, \label{eq:msumpos}
\end{align}
\end{subequations}
The antisymmetry of \(M\) gives \eqref{eq:msum} and \eqref{eq:msumpos}, \eqref{eq:deltapjsum} follows from the definition of the \(\Delta p_{j}\), and \eqref{eq:jdeltapjsum} follows from \eqref{eq:msum} and \eqref{eq:deltapjsum}. Because \(\Gprod\) is self dual \(\overline{X} \subset \Gprod \iff X \subset G^{\otimes k}_{N}\), which will save us from having to take the conjugate of \(X\) in \eqref{eq:trivialcondition}. Writing the highest weight (Dynkin label) of \(X\) as \(\qty(p_{1}, p_{2}, \ldots, p_{N - 1})\) we see that
\begin{equation}
    X \subset \Gprod \implies p_{j} = \delta^{1}_{j} k + \delta^{N-1}_{j} k - \Delta p_{j},
\end{equation}
for some \(\Delta p\). Using \eqref{eq:jdeltapjsum} we find that
\begin{equation} \label{eq:jpjsum}
    \sum^{N-1}_{j = 1} j p_{j} = N \qty(k + m_{N}).
\end{equation}
Because \(m_{N}\) is integral the r.h.s. of \eqref{eq:jpjsum} is some integer multiple of \(N\). In order for a \(\kmin\) to exist the l.h.s. of \eqref{eq:jpjsum} must then also be some integer multiple of \(N\):
\begin{equation} \label{eq:pjmodN}
    \exists \, k \in \mathbb{Z}^{+} \mid \mathbbm{1} \subset X \otimes \Gprod \iff \sum^{N-1}_{j = 1} j p_{j} = 0 \mod N.
\end{equation}
Notice that the right-hand side of the expression above corresponds to the vanishing \(N\)-ality condition.

Next we sum over the \(p_{j}\) from \(n \leq j \leq N-1\) and use \eqref{eq:deltapjsum} to solve for the \(m_{j}\):
\begin{equation}
\begin{split}
    m_{n} =& \; m_{N} + \sum^{N-1}_{j = n} \qty(\delta^{1}_{j} + \delta^{N-1}_{j})k - \sum^{N-1}_{j = n} p_{j}, \quad n < N,\\[2mm]
    m_{N} =& \; \frac{1}{N} \sum^{N-1}_{j = 1} j p_{j} - k.
\end{split}
\end{equation}
Substituting these values of \(m_{j}\) into \eqref{eq:msumpos} then gives the requirement
\begin{equation} \label{eq:kn}
    k \geq \frac{N - n}{N} \sum^{N-1}_{j = 1} j p_{j} - \sum^{N-1}_{j = n + 1} \qty(j - n) p_{j}, \quad 1 \leq n \leq N - 1.
\end{equation}
To simplify notation define
\begin{equation} \label{eq:kndef}
    k_{n} \coloneqq \frac{n}{N} \sum^{N-1}_{j = 1} j p_{j} - \sum^{n}_{j = 1} \qty(n - j) p_{N - j}
\end{equation}
The largest of these \(k_{n}\) then satisfies \eqref{eq:kn} for all \(n\), and is the smallest such value of \(k\) that does so. We then have the final result:
\begin{equation} \label{eq:kmin}
    k_{\mathrm{min}} = \max\qty{k_{n} \mid 1 \leq n \leq N - 1}. 
\end{equation}

\subsection{Including Matter Fields}\label{sec:matter}

We next include an additional matter field in the tensor product
\begin{equation*}
    X \otimes Q_{i} \otimes \Gprod,
\end{equation*}
where \(X\) is some field in an arbitrary representation of \(\SU(N)\), \(G_{N}\) is again a gauge field in the adjoint representation, and \(Q_{i}\) is a matter field in the
\begin{equation}
    \dim(Q_{i}) = \frac{N!}{i ! \qty(N - i)!}, \quad 1 \leq i \leq N - 1,
\end{equation}
representation. The Dynkin label for \(Q_{i}\) is
\begin{equation}
\begin{split}
    Q_{i} \cong&\; (\underbrace{0, \ldots, 0}_{i - 1}, 1, \underbrace{0, \ldots, 0}_{N - i - 1}),\\[2mm]
    \cong& \; \vcenter{\hbox{\begin{tikzpicture}
    \node[Young tableau](yt){ \\
     \\
     |[vdots]| \\
     \\
    };
    \draw[very thick, decoration={calligraphic brace, amplitude=6pt}, decorate] (yt-1-1.north east) -- (yt-4-1.south east) node[midway, above=1ex, sloped]{\(i\)};
    \end{tikzpicture}}}
\end{split}
\end{equation}
If the \irrep{} \(X\) has highest weight \(p\) then the highest weight of the highest \irrep{} in the tensor product \(X \otimes Q_{i}\) is
\begin{equation*}
    \qty(p_{1}, \ldots, p_{i - 1}, p_{i} + 1, p_{i + 1}, \ldots, p_{N - 1}).
\end{equation*}
Only one \irrep{} in the tensor product can have this weight in its weight space. The other \irreps{} can again be built using a similar procedure to the previous section. Begin with the Dynkin label for the highest \irrep{} above and move boxes to lower rows in all possible ways, keeping the vertical ordering of the boxes intact. All of the resulting Dynkin labels take the form
\begin{equation}
\begin{split}
    f_{1} =& \; p_{1} + \delta^{i}_{1} - 2 i_{12} + i_{23}\\[2mm]
    f_{j} =& \; p_{j} + \delta^{i}_{j} + i_{j - 1 \; j} - 2 i_{j \; j+1} + i_{j + 1 \; j + 2},\\[2mm]
    f_{N-1} =& \; p_{N-1} + \delta^{i}_{N - 1} - 2 i_{N - 1 \; N} + i_{N - 2 \; N - 1},
\end{split}
\end{equation}
where the integers \(i_{jk}\) correspond to moving \(i_{jk}\) boxes from row \(j\) to row \(k\). Again, we compact this procedure into an \(N \times N\) matrix:
\begin{equation}
    M \equiv \mqty*(0 & i_{12} & 0 & 0 & \ldots & 0 \\
    - i_{12} & 0 & i_{23} & 0 & \ldots & 0 \\
    \vdots & \vdots & \vdots & \vdots & & \vdots\\
    & - i_{i - 1 \; i} & 0 & i_{i \: i + 1} & 0 & \\
    & 0 & - i_{i \; i + 1} & 1 & i_{i+1 \; i+2} & \\
    \vdots & \vdots & \vdots & \vdots & \ddots & \vdots\\
    0 & \ldots & \ldots & 0 & - i_{N - 1 \; N} & 1).
\end{equation}
Then, as in the previous section, define
\begin{subequations}
\begin{align}
    m_{j} =& \; \sum^{N}_{k = 1} M_{j k},\\[2mm]
    \Delta p_{j} =& \; m_{j} - m_{j + 1}.
\end{align}
\end{subequations}
The Dynkin labels for the resulting \irreps{} are then
\begin{equation} \label{eq:fdeltap}
    f_{j} = p_{j} - \Delta p_{j}.
\end{equation}
We then substitute \eqref{eq:fdeltap} into \eqref{eq:kndef} in preparation for determining \(\kmin\):
\begin{multline} \label{eq:knwithmatter}
    k_{n} = \frac{n}{N} \sum^{N-1}_{j = 1} j p_{j} - \sum^{n}_{j = 1} \qty(n - j) p_{N - j}\\[2mm]
    + \sum^{N}_{j = N - n + 1} m_{j}  - \frac{\qty(N - i) n}{N}.
\end{multline}
This gives a \(\kmin\) for every unique set of \(f_{j}\). There will exist at least one set of \(f_{j}\) that gives the smallest \(\kmin\), but to determine which set we must consider the restrictions on the \irreps{} in \(X \otimes Q_{i}\).

There is a condition on the value of \(i\) for a \(\kmin\) to exist, because we must have
\begin{equation}
    \sum^{N-1}_{j = 1} j f_{j} = 0 \mod{N}
\end{equation}
from \eqref{eq:pjmodN}. Since \(1 \leq i \leq N - 1\) it must be fixed so that
\begin{equation}
    i = N - \qty(\sum^{N-1}_{j = 1} j p_{j} \mod{N}).
\end{equation}
Because \(Q_{i}\) corresponds to an antisymmetric combination of fundamental indicies\footnote{This \irrep{} is generated by the Schur functor \(\mathbb{S}^{(1, \ldots, 1)} V = \wedge^{(1, \ldots, 1)} V\), where \(V = \mathbb{C}^{N}\), hence the antisymmetry.} (for \(i > 1\)) we can place no more than a single box in each row. As well, we can move at most one box to the very bottom of the tableau (and remove the completed column). This implies \(i_{N -1 \; N} = 0\) or \(1\), which in turn implies \(m_{N} = 0\) or \(1\). If no boxes are moved to the bottom of the tableau, i.e. \(i_{N - 1 \; N} = 0\), then every row can have, at most, one more box than the same row in the original tableau for \(X\). In addition, every row must have at least as many boxes than the same row in \(X\). If \(i_{N -1 \; N} = 1\) then we eliminate the completed column, thus shortening each row by one box. In this case every row can contain \textit{no more} boxes than the same row in \(X\) and \textit{no fewer} than one less than the number of boxes in the same row in \(X\). In terms of the Dynkin labels \(f\) this is
\begin{equation} \label{eq:fconditions}
    \sum^{N-1}_{j = k} p_{j} - i_{N - 1 \; N} \leq \sum^{N-1}_{j = k} f_{j} \leq 1 - i_{N - 1 \; N} + \sum^{N - 1}_{j = k} p_{j}.
\end{equation}
Inserting \eqref{eq:fdeltap} into \eqref{eq:fconditions} then yields the following conditions on the \(m_{j}\)
\begin{equation} \label{eq:mcond1}
    0 \leq m_{j} \leq 1.
\end{equation}
Summing over all of the \(m_{j}\) gives
\begin{equation} \label{eq:mcond2}
    \sum^{N}_{j = 1} m_{j} = N - i.
\end{equation}
And, finally, requiring all of the \(f_{j} \geq 0\) gives
\begin{equation} \label{eq:mcond3}
    m_{j} - m_{j + 1} \leq p_{j}.
\end{equation}

The only free parameters in the \(k_{n}\) are the \(m_{j}\). In order to find the smallest possible \(\kmin\) we attempt to minimize as many of the \(k_{n}\) as possible. This implies looking for the set of \(m_{j}\), satisfying \eqref{eq:mcond1}, \eqref{eq:mcond2}, and \eqref{eq:mcond3}, such that the sum \(\sum^{N}_{j = N - n + 1} m_{j}\) is as small as possible for as many \(n\) as possible. The best case scenario is
\begin{equation} \label{eq:bestcase}
\begin{split}
    m_{1} = m_{2} = \ldots&\; = m_{N - i} = 1.\\[2mm]
    m_{N - i + 1} = m_{N - i + 2} = \ldots&\; = m_{N} = 0
\end{split}
\end{equation}
However, we cannot guarantee that such a configuration satisfies all of the constraints for arbitrary \(X\). In particular, for the configuration in \eqref{eq:bestcase} the condition \eqref{eq:mcond3} gives
\begin{equation}
    m_{N - i} - m_{N - i + 1} = 1 \implies p_{N - i} \geq 1,
\end{equation}
which is clearly only a special case. We must have \(N - i\) of the \(m_{j}\) equal to 1 from \eqref{eq:mcond1} and \eqref{eq:mcond2}, but we would prefer to have the non-zero \(m_{j}\) occur mostly when \(j\) is small. This corresponds to moving as many of the \(i\) boxes to the bottom of the tableau as possible. To this end define the set
\begin{equation}
    L \coloneqq \qty{ l \mid p_{l} \neq 0} \cup \qty{0, N}.
\end{equation}
Note that, with the exception of the trivial representation, this set contains at least 3 elements for any \irrep{}. Equivalently, we can embed the \irrep{} \(X\) in a larger space:
\begin{equation}
\begin{split}
    \tilde{X} \cong& \; \qty(\tilde{p}_{0}, \tilde{p}_{1}, \ldots, \tilde{p}_{N - 1}, \tilde{p}_{N}),\\[2mm]
    \cong& \; (1, \underbrace{p_{1}, \ldots, p_{N - 1}}_{X}, 1).
\end{split}
\end{equation}
The set \(L\) is then
\begin{equation}
    L = \qty{ l \mid \tilde{p}_{l} \neq 0}.
\end{equation}
Then find the largest \(l\) in this set such that \(l \leq N - i\):
\begin{equation}
    \ell_{1} = \max\qty{l \mid l \in L, l \leq N - i}.
\end{equation}
If \(\ell_{1} \neq 0\) we can always let \(m_{\ell_{1}} = 1\), since \(p_{\ell_{1}} > 0\). Next, we look for a second \(l \in L\) such that it is the minimum value satisfying \(l > \ell_{1}\).
\begin{equation}
    \ell_{2} = \min\qty{l \mid l \in L, l > \ell_{1}}.
\end{equation}
Provided \(\ell_{2} \neq N\) we can always allow \(m_{\ell_{2} + 1} = 0\) because \(p_{\ell_{2}} > 0\). Note that because \(\ell_{2} > N - i\) not all of the \(i\) boxes coming from \(Q_{i}\) can be placed below row \(\ell_{2}\). However, we can always place \(N - \ell_{2}\) boxes below row \(\ell_{2}\). The remaining \(i - N + \ell_{2}\) boxes can then be placed directly below row \(\ell_{1}\). This moves as many boxes as possible to the bottom of the tableau.

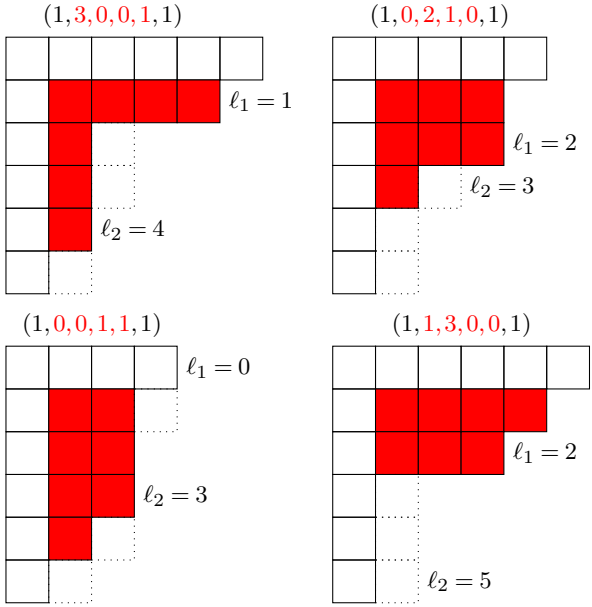
\begin{figure}
    \noindent\begin{minipage}[t]{0.24\textwidth}\flushleft%
        \begin{tikzpicture}
            \node[Young tableau](yt){ & & & & & \\
             & |[fill=red]| & |[fill=red]| & |[fill=red]| & |[fill=red]| \\
             & |[fill=red]| & |[dotted]| \\
             & |[fill=red]| & |[dotted]| \\
             & |[fill=red]| \\
             & |[dotted]| \\
             };
            \node[above] at (yt-1-3.north) {\(\qty(1, {\color{red} 3, 0, 0, 1}, 1)\)};
            \node[right] at (yt-2-5.east) {\(\ell_{1} = 1\)};
            \node[right] at (yt-5-2.east) {\(\ell_{2} = 4\)};
        \end{tikzpicture}\hfill
    \end{minipage}\hfill%
    \begin{minipage}[t]{0.24\textwidth}\flushleft%
        \begin{tikzpicture}%
            \node[Young tableau](yt){ & & & & \\
             & |[fill=red]| & |[fill=red]| & |[fill=red]| \\
             & |[fill=red]| & |[fill=red]| & |[fill=red]| \\
             & |[fill=red]| & |[dotted]| \\
             & |[dotted]| \\
             & |[dotted]| \\
             };
            \node[above] at (yt-1-3.north) {\(\qty(1, {\color{red} 0, 2, 1, 0}, 1)\)};
            \node[right] at (yt-3-4.east) {\(\ell_{1} = 2\)};
            \node[right] at (yt-4-3.east) {\(\ell_{2} = 3\)};
        \end{tikzpicture}\hfill
    \end{minipage}
    
    \noindent\begin{minipage}[t]{0.24\textwidth}\flushleft%
        \begin{tikzpicture}%
            \node[Young tableau](yt){ & & & \\
             & |[fill=red]| & |[fill=red]| & |[dotted]| \\
             & |[fill=red]| & |[fill=red]| \\
             & |[fill=red]| & |[fill=red]| \\
             & |[fill=red]| & |[dotted]| \\
             & |[dotted]| \\
             };
            \node[above] at (yt-1-2.north east) {\(\qty(1, {\color{red} 0, 0, 1, 1}, 1)\)};
            \node[right] at (yt-1-4.east) {\(\ell_{1} = 0\)};
            \node[right] at (yt-4-3.east) {\(\ell_{2} = 3\)};
        \end{tikzpicture}\hfill
    \end{minipage}\hfill%
    \begin{minipage}[t]{0.24\textwidth}\flushleft%
        \begin{tikzpicture}
            \node[Young tableau](yt){ & & & & & \\
             & |[fill=red]| & |[fill=red]| & |[fill=red]| & |[fill=red]| \\
             & |[fill=red]| & |[fill=red]| & |[fill=red]| \\
             & |[dotted]| \\
             & |[dotted]| \\
             & |[dotted]| \\
             };
            \node[above] at (yt-1-3.north east) {\(\qty(1, {\color{red} 1, 3, 0, 0}, 1)\)};
            \node[right] at (yt-3-4.east) {\(\ell_{1} = 2\)};
            \node[right] at (yt-6-2.east) {\(\ell_{2} = 5\)};
        \end{tikzpicture}\hfill
    \end{minipage}%
    \caption{Four examples for determining the optimum \irrep{} in \(X \otimes Q_{i}\) with \(N = 5\) and \(i = 3\). Shown in red is the original \irrep{} \(X\), embedded inside the larger space \(\tilde{X}\). The dotted boxes show the placement of boxes that guarantees the smallest possible \(k_{\mathrm{min}}\). From left to right, the resulting minimum number of copies of the adjoint representation are \(k_{\mathrm{min}} = 2, 2, 1, \text{ and } 3\). \label{fig:matterfieldexamples}}
\end{figure}

Now we consider how the procedure we have just outlined changes the Dynkin label of \(\tilde{X}\). Placing \(i - N + \ell_{2}\) boxes below row \(\ell_{1}\) will reduce \(\tilde{p}_{\ell_{1}}\) by 1 and increase the \(\tilde{p}_{i - N + \ell_{1} + \ell_{2}}\) by 1. Similarly, placing \(N - \ell_{2}\) boxes below row \(\ell_{2}\) reduces \(\tilde{p}_{\ell_{2}}\) by one and increases \(\tilde{p}_{N}\) by 1. Therefore, define
\begin{equation} \label{eq:deltatildep}
    \Delta \tilde{p}_{j} = \delta^{\ell_{1}}_{j} - \delta^{i - N + \ell_{1} + \ell_{2}}_{j} + \delta^{\ell_{2}}_{j} - \delta^{N}_{j},
\end{equation}
where \(j\) runs from 0 to \(N\), so that the Dynkin labels in the extended space become
\begin{equation}
    \tilde{f}_{j} = \tilde{p}_{j} - \Delta \tilde{p}_{j},
\end{equation}
Four examples have been given in Fig.~\ref{fig:matterfieldexamples}. Note that if \(\ell_{1} = N - i\) the second and third term on the r.h.s. of \eqref{eq:deltatildep} cancel. This is because \(\ell_{2}\) becomes irrelevant in this case since all of the \(i\) boxes can be placed below row \(\ell_{1}\), creating a complete column. If \(\ell_{2} = N\) it may not be possible to complete a column, and the last two terms on the r.h.s. of \eqref{eq:deltatildep} cancel as expected.

Restricting \(j\) to values greater than 0 and less than \(N\) gives
\begin{equation} \label{eq:fjdelta}
    \begin{split}
    f_{j} =& \; \tilde{f}_{j}, \quad 1 \leq j \leq N - 1,\\[2mm]
    =& p_{j} - \delta^{\ell_{1}}_{j} + \delta^{i - N + \ell_{1} - \ell_{2}}_{j} - \delta^{\ell_{2}}_{j}.
    \end{split}
\end{equation}
We can then substitute \eqref{eq:fjdelta} directly into \eqref{eq:kndef}, yielding
\begin{multline}
    k_{n} = \frac{n}{N} \sum^{N -1}_{j = 1} j p_{j} - \sum^{n}_{j = 1} \qty(n - j) p_{N - j}\\[2mm]
    - \frac{n}{N} \sum^{N - 1}_{j = 1} j \qty(\delta^{\ell_{1}}_{j} - \delta^{i - N + \ell_{1} + \ell_{2}}_{j} + \delta^{\ell_{2}}_{j} - \delta^{N}_{j})\\[2mm]
    + \sum^{n}_{j = 1} \qty(n - j) \qty(\delta^{\ell_{1}}_{N - j} - \delta^{i - N + \ell_{1} + \ell_{2}}_{N - j} + \delta^{\ell_{2}}_{N - j} - \delta^{N}_{N - j}).
\end{multline}
This can be simplified using
\begin{multline}
    \sum^{N - 1}_{j = 1} j \qty(\delta^{\ell_{1}}_{j} - \delta^{i - N + \ell_{1} + \ell_{2}}_{j} + \delta^{\ell_{2}}_{j} - \delta^{N}_{j})\\[2mm]
    + N \delta^{\ell_{1}}_{N} - N \delta^{i - N + \ell_{1} + \ell_{2}}_{N} + N \delta^{\ell_{2}}_{N} - N\\[2mm]
    = \sum^{N}_{j = 0} j \qty(\delta^{\ell_{1}}_{j} - \delta^{i - N + \ell_{1} + \ell_{2}}_{j} + \delta^{\ell_{2}}_{j} - \delta^{N}_{j}) = - i, 
\end{multline}
and
\begin{multline}
    \sum^{n}_{j = 1} \qty(n - j) \qty(\delta^{\ell_{1}}_{N - j} - \delta^{i - N + \ell_{1} + \ell_{2}}_{N - j} + \delta^{\ell_{2}}_{N - j} - \delta^{N}_{N - j})\\[2mm]
    + n \delta^{\ell_{1}}_{N} - n \delta^{i - N + \ell_{1} + \ell_{2}}_{N} + n \delta^{\ell_{2}}_{N} - n\\[2mm]
    = - i\\[2mm]
    + \sum^{N}_{j = n + 1} \qty(n - j) \qty(\delta^{\ell_{1}}_{N - j} - \delta^{i - N + \ell_{1} + \ell_{2}}_{N - j} + \delta^{\ell_{2}}_{N - j} - \delta^{N}_{N - j})
\end{multline}
After this simplification we have
\begin{multline} \label{eq:knmatter}
    k_{n} = \frac{n}{N} \sum^{N -1}_{j = 1} j p_{j} - \sum^{n}_{j = 1} \qty(n - j) p_{N - j} - \frac{\qty(N - n) i}{N}\\[2mm]
    - \sum^{N - n - 1}_{j = 0} \qty(N - n - j) \qty(\delta^{\ell_{1}}_{j} - \delta^{i - N + \ell_{1} + \ell_{2}}_{j} + \delta^{\ell_{2}}_{j}),
\end{multline}
and our final result is, as before, the maximum of these \(k_{n}\):
\begin{equation}\label{eq:finalres}
    \kmin = \max\qty{ k_{n} \mid 1 \leq n \leq N - 1}.
\end{equation}

Especially for large $N$, evaluating Eq.~(\ref{eq:finalres}) becomes computationally intensive and highly non-trivial. We therefore built and made publicly available the {\tt tessellation} python code. We describe how to install and use the code in App.~\ref{app:code}. 

\normalsize
\section{The lowest dimensional portals}\label{sec:code}
Consider an exotic species belonging to an irreducible representation $X$ of SU($N$). Assume that the quantum field associated with $X$ has mass-dimension $d_X$, and that matter fields have mass-dimension $d_q$ (for instance $d_q=3/2$ for fermions, such as quarks or antiquarks) and that gauge-mediator fields in the adjoint representation have mass dimensions $d_g$ (for instance, $d_g=1$ for gluons). We are interested in gauge-invariant ``portal'' operators, with the structure
\begin{equation}\label{eq:operator}
    {\cal O}_j=\frac{c_j}{\Lambda^{n_j}}X\left(\prod_{i=1}^{N-1} Q_i^{n_{Q_i}}\right)g^{n_g},
\end{equation}
where $Q_i$ represents the quantum field corresponding to one of the $N-1$ fundamental representations of SU($N$), with Dynkin label $$Q_i \cong (\underbrace{0, \ldots, 0}_{i - 1}, 1, \underbrace{0, \ldots, 0}_{N - i - 1}),$$
and where $g$ is the quantum field corresponding to the gauge boson of SU($N$), with Dynkin label (1,0,\dots,0,1). Note that we assume 4 space-time dimensions, so the mass dimensions of the operator in Eq.~(\ref{eq:operator}) must equal 4, and we assume that $\Lambda$ has mass dimension 1, as customary. Also, we assume that the spin of $X$ is such that ${\cal O}_j$ respects Poincar\'e invariance.

Here, we seek the lowest possible value of $n_j$ such that the operator in Eq.~(\ref{eq:operator}) is gauge invariant, i.e. the lowest dimensional ``portal'' for $X$. Notice that the condition that the operator be gauge invariant is equivalent to the condition that the tensor product of the representations is such that
\begin{equation}
    \mathbbm{1} \subset X\otimes \bigotimes_{i=1}^{N-1} Q_i^{\otimes n_{Q_i}}\otimes G_N^{\otimes n_g}.
\end{equation}
The results above address precisely this question: given the $N$-ality of $X$, say $t$, $n_i=0$ $\forall i$ if $t=0$, and $n_{Q_{N-t}}=1$, $n_i=0$ $\forall i\neq N-t$ for $t\neq0$, while $n_g=k_{\rm min}$ with $k_{\rm min}$ given in Eq.~(\ref{eq:finalres}). As a result, we find
\begin{equation}
\begin{split}
    4&=-n_j+d_X+d_q+k_{\rm min}d_g\\
    &\Downarrow\\
    n_j&=d_X+d_q+k_{\rm min}d_g-4.
\end{split}
\end{equation}
For instance, let's consider the QCD case in the Standard Model, with gauge group SU(3); let's assume $X\cong (p,q)$, thus $t=(p+2q)\ {\rm mod}\ 3$. We have the following cases:
\begin{enumerate}[(i)]
    \item \(t = 0\): this implies \(n_{Q_{i}} = 0\) (\(i = 1, 2\)),
    \begin{equation*}
        \kmin = \begin{cases} \frac{p + 2 q}{3} & p \leq q \\ \frac{2 p + q}{3} & p \geq q \end{cases},
    \end{equation*}
    and \(n_{j} = d_{X} + \kmin d_{g} - 4\).
    \item \(t = 1\): this implies \(n_{Q_{1}} = 0\), \(n_{Q_{2}} = 1\),
    \begin{equation*}
        \kmin = \begin{cases} \frac{p + 2 q - 1}{3} & p - 1 \leq q \\ \frac{2 p + q - 2}{3} & p - 1 \geq q \end{cases},
    \end{equation*}
    and \(n_{j} = d_{X} + d_{q} + \kmin d_{g} - 4\).
    \item \(t = 3\): this implies \(n_{Q_{1}} = 1\), \(n_{Q_{2}} = 0\),
    \begin{equation*}
        \kmin = \begin{cases} \frac{p + 2 q - 2}{3} & p \leq q - 1 \\ \frac{2 p + q - 1}{3} & p \geq q - 1 \end{cases},
    \end{equation*}
    and \(n_{j} = d_{X} + d_{q} + \kmin d_{g} - 4\).
\end{enumerate}
Similar cases for $N>3$ and a different matter content can be readily computed utilizing the companion ${\tt tessellation}$ code.

\section{Summary and conclusions}\label{sec:conclusions}
In this study we have entertained the question of which is the lowest-possible mass dimension of SU($N$) gauge invariant operators that contain an exotic new state in a given representation $X$ of SU($N$). We showed that to address this question one needs to find which is the smallest tensor product of the adjoint representation that contains the representation $X$. We fully solved this question, and built a numerical code that solves this problem for any $N$. We then solved the original question, pointing out that one needs to first take the tensor product of $X$ with one of the fundamental representations of SU($N$) to obtain a representation with vanishing $N$-ality, and then find the minimal number of copies of the adjoint representation that contains that product. We gave explicit formulae for any $N$, and explicit results for the mass dimension of the lowest-dimensional portal for the case of SU(3) of QCD. Our results are relevant for several areas of particle phenomenology which seek new physics beyond the Standard Model: the numerical tool {\tt tessellation} is also per se relevant in the context of group theory.

\begin{acknowledgments}
We gratefully acknowledge helpful conversations with Martin Weissman. SP and JU are partly supported by the U.S.\ Department of Energy grant number de-sc0010107.
\end{acknowledgments}

\appendix

\section{Roots, Weights, and Representations} \label{app:representations}

In this appendix we briefly cover some basic concepts in representation theory and a few common representations of \(\SU(N)\). A representation \(\rho\) of a group \(G\) is a linear mapping of the elements of the group to the automorphisms of some vector space \(V\) that preserves the group multiplication.
\begin{equation}
\begin{split}
    \rho \colon G \to& \; \GL(V),\\[2mm]
    g \mapsto& \; \rho(g), \quad g \in G,\\[2mm]
    \rho(g) \cdot \rho(h) =& \; \rho(g \circ h) \quad \forall g, h \in G,
\end{split}
\end{equation}
where \(\circ\) is the multiplication on \(G\) and \(\cdot\) is the multiplication on \(\GL(V)\). If the representation space \(V\) is finite dimensional then \(\cdot\) is just matrix multiplication between \(\dim(V) \times \dim(V)\) matrices.

The fundamental representation of \(\SU(N)\) is defined by the mapping
\begin{equation}
    \rho_{F} \colon \SU(N) \to \GL(\mathbb{C}^{N}),
\end{equation}
i.e. we choose \(V = \mathbb{C}^{N}\). We can then choose a basis \(\qty{\vb{e}_{a}}\) on \(\mathbb{C}^{N}\):
\begin{equation}
    \psi = T^{a} \vb{e}_{a}, \quad T^{a} \in \mathbb{C}, \forall \psi \in V.
\end{equation}
The \textit{dual} vector space to \(V\), which we denote \(V^{*}\), is the vector space consisting of linear functionals of \(V\) to the base field of \(V\).
\begin{equation}
    V^{*} \colon V \to K,
\end{equation}
where \(K\) is the field underlying the vector space \(V\). A choice of basis on \(V\) induces a natural basis on \(V^{*}\), 
\begin{equation}
    \xi = T_{a} \vb{e}^{a}, \quad T_{a} \in \mathbb{C}, \; \forall \xi \in V^{*}.
\end{equation}
where the \(\qty{\vb{e}^{a}}\) form the \textit{dual basis of the dual space}, i.e. \(\vb{e}^{b} \qty(\vb{e}_{a}) = \delta^{b}_{a}\). Because \(V\) is finite dimensional \(V^{*}\) is isomorphic to \(V\) as a vector space. The adjoint representation of \(\SU(N)\), denoted \(\Ad\), is a map from the group into the automorphisms of the associated Lie algebra:
\begin{equation}
\begin{split}
    \Ad \colon \SU(N) \to& \operatorname{Aut}(\su(N))\\[2mm]
    g \mapsto& \Ad_{g}, \quad g \in \SU(N).
\end{split}
\end{equation}
The group action on a vector in the adjoint representation is
\begin{equation}
    \mathrm{Ad}_{g} (\eta) = g \eta g^{-1}, \quad \eta \in \mathfrak{su}(N). 
\end{equation}
Because \(SU(N)\) is a closed subgroup of the general linear group the above hold for all \(g \in SU(N)\) and all \(\eta \in \mathfrak{su}(N)\). Thus, we have the basis for adjoint representation
\begin{equation}
    A = T^{a}_{b} \vb{e}_{a} \otimes \vb{e}^{b}, \quad A \in \mathfrak{su}(N), \; T^{a}_{b} \in \mathbb{R},
\end{equation}
with
\begin{equation}
    T^{a}_{a} = 0.
\end{equation}
We have introduced the tensor product notation, which we now explain. A tensor is a generalization of the notion of the dual vector space. It is a multilinear functional that takes some copies of the vector spaces \(V\) and \(V^{*}\) and maps them to the base field.
\begin{equation}
    T^{r}_{s} V \colon \underbrace{V \times \ldots \times V}_{r} \times \underbrace{V^{*} \times \ldots \times V^{*}}_{s} \to K,
\end{equation}
where \(K\) is the base field underlying the vector space \(V\). With the previous choice of basis we have
\begin{equation}
\begin{split}
    T &= T^{a_{1} \ldots a_{r}}_{b_{1} \ldots b_{s}} \vb{e}_{a_{1}} \otimes \ldots \otimes \vb{e}_{a_{r}} \otimes \vb{e}^{b_{1}} \otimes \ldots \otimes \vb{e}^{b_{s}},\\[2mm]
    &\phantom{=} T \in T^{r}_{s} V.
\end{split}
\end{equation}
We can inherit the obvious addition and scalar multiplication induced by \(V\), making \(T^{r}_{s} V\) a \(\dim(V)^{r + s}\) vector space over the field \(K\). It is then clear that the \(\vb{e}_{a_{1}} \otimes \ldots \otimes \vb{e}_{a_{r}} \otimes \vb{e}^{b_{1}} \otimes \ldots \otimes \vb{e}^{b_{2}}\) constitutes a basis on \(T^{r}_{s} V\). This can be generalized to the tensor product of different vector spaces \(V \otimes W\). A vector \(\psi \otimes \xi\), where \(\psi \in V\) and \(\xi \in W\), is then an element of \(V \otimes W\). Generally the space \(V \otimes W\) is much larger than the space spanned by vectors of the form \(\psi \otimes \xi\), it will also include linear combinations of vectors. As well, we must take \(\psi \otimes \xi\) to represent equivalence classes, because there will be multiple \(\psi\) and \(\xi\) that yield the same vector in \(V \otimes W\). Also note that we use the same symbol, \(\otimes\), to represent the tensor product between \textit{vector spaces} and \textit{vectors}, as is common in the literature. They are in fact different notions of a product, but the usage is clear from context.

For a complex semisimple Lie algebra \(\mathfrak{g}\) with a Cartan subalgebra \(\mathfrak{h}\) the \textit{root vectors} are elements of \(\mathfrak{h}^{*}\), i.e. the dual of the Cartan subalgebra. A \(\alpha \in \mathfrak{h}^{*}\) is called a root of \(\mathfrak{g}\) relative to \(\mathfrak{h}\) if \(\alpha \neq 0\) and there exists an \(X \neq 0 \in \mathfrak{g}\) such that
\begin{equation} \label{eq:rootequation}
    \qty[H, X] = \alpha (H) X, \quad \forall H \in \mathfrak{h}.
\end{equation}
The Cartan subalgebra of \(\su(N)\) has \(N-1\) generators that commute with all other elements of the algebra. Hence, \(\mathfrak{h}\) is an \(N-1\) dimensional vector space over \(\mathbb{C}\) (in the complexified version of \(\su(N)\)). It is, however, somewhat simpler to consider the generators of \(\mathfrak{h}\) as elements of \(\mathbb{R}^{N}\) with the restriction that the sum over all \(N\) components vanishes, that is
\begin{equation}
    H_{\lambda} = \lambda^{j} \vb{e}_{j}, \quad \sum^{N}_{j = 1} \lambda^{j} = 0,
\end{equation}
where the \(\qty{\vb{e}_{j}}\) are the usual orthonormal basis vectors on \(\mathbb{R}^{N}\). Since the roots \(\alpha\) lie in the dual space to \(\mathfrak{h}\), i.e. \(\mathfrak{h}^{*}\), we can use the dual basis to write them as
\begin{equation}
    \alpha^{j k} = \vb{e}^{j} - \vb{e}^{k}, \quad \vb{e}^{j}(H_{\lambda}) = \lambda^{j}, \quad j \neq k.
\end{equation}
We then have
\begin{equation}
    \qty[H_{\lambda}, E_{jk}] = \qty(\lambda^{j} - \lambda^{k}) E_{j k},
\end{equation}
where the \(E_{jk}\) are \(N \times N\) matrices with a 1 in the \(\qty(j,k)\)\textsuperscript{th} position and zeros elsewhere in the fundamental representation (there is no sum over \(j\) and \(k\)). The simple roots of the algebra are given by
\begin{equation}
    \alpha^{j} \coloneqq \alpha^{j \, j + 1}.
\end{equation}
All of the other roots can be constructed by linear combinations of the simple roots. This root system is known as \(A_{N-1}\), and we will denote the set of all roots as \(\Delta\). The simple roots are also vectors in \(\mathbb{R}^{N}\) whose components sum to 0 due to the isomorphism between a finite dimensional vector space and its dual. This is the \textit{standard basis} (sometimes called the \textit{orthogonal-basis}:
\begin{equation}
\begin{split}
    \alpha^{1} =& \; (1, - 1, \underbrace{0, \ldots, 0}_{N-2}),\\[2mm]
    \alpha^{j} =& \; (\underbrace{0, \ldots, 0}_{j-1}, 1, -1, \underbrace{0, \ldots, 0}_{N - j - 1}),\\[2mm]
    \alpha^{N-1} =& \; (\underbrace{0, \ldots, 0}_{N - 2}, 1, -1).
\end{split}
\end{equation}
The simple roots form a complete basis for \(\mathbb{R}^{N-1}\), but they are not orthonormal. We can therefore express any vector in \(\mathbb{R}^{N-1}\) as a linear combination of the simple roots
\begin{equation}
    v \in \mathbb{R}^{N-1} = v_{i} \alpha^{i}, \quad v_{i} \in \mathbb{R}.
\end{equation}
This is referred to as the \textit{\(\alpha\)-basis}. It is also useful to construct the \textit{fundamental weights} \(\omega^{j}\), defined by
\begin{equation}
    \frac{2 \expval{\omega^{j}, \alpha^{k}}}{\expval{\alpha^{k}, \alpha^{k}}} = \delta^{k}_{j}
\end{equation}
where \(\expval{\cdot, \cdot}\) is the usual inner product on \(\mathbb{R}^{N}\). The \(\omega^{j}\) constitute the \textit{\(\omega\)-basis}, or \textit{Dynkin-basis}. We can translate between the \(\alpha\)-basis and \(\omega\)-basis via
\begin{subequations}
\begin{align}
    \alpha^{j} =& \; \sum^{N - 1}_{k = 1} A^{j}_{k} \omega^{k},\\[2mm]
    \omega^{j} =& \; \sum^{N-1}_{k = 1} \qty(A^{-1})^{j}_{k} \alpha^{k},
\end{align}
\end{subequations}
where the \textit{Cartan matrix} \(A\) is simply
\begin{equation}
    A^{j}_{k} = \frac{2 \expval{\alpha^{k}, \alpha^{j}}}{\expval{\alpha^{k}, \alpha^{k}}}.
\end{equation}
As previously stated, for \(\SU(N)\) the root system is \(A_{N-1}\), and the Cartan matrix is
\begin{equation}
    A^{j}_{k} = \mqty*(2 & -1 & 0 & \ldots & 0 & 0\\
    -1 & 2 & -1 & \ldots & 0 & 0\\
    0 & -1 & 2 & \ldots & 0 & 0\\
    \vdots & \vdots & \vdots &  \ddots & \vdots & \vdots\\
    0 & 0 & 0 & \ldots & 2 & -1\\
    0 & 0 & 0 & \ldots & - 1 & 2).
\end{equation}
We can make the two basis dual by introducing the \textit{coroot} \(\alpha^{\vee j k}\):
\begin{equation}
    \alpha^{\vee}_{j k} \coloneqq \frac{2 \alpha^{j k}}{\expval{\alpha^{j k}, \alpha^{j k}}}.
\end{equation}
This then gives
\begin{equation}
    \expval{\omega^{j}, \alpha^{\vee}_{k}} = \delta^{j}_{k},
\end{equation}
where \(\alpha^{\vee}_{j}\) is the coroot associated to the simple root \(\alpha^{j}\).

It then follows that a representation of a Lie algebra is a set of matrices that satisfy the same commutation relations as the abstract algebra. Each of the matrices can be labeled by a \textit{weight} as follows: Let \(V\) be a representation of a Lie algebra \(\mathfrak{g}\) over \(\mathbb{C}\) and let \(\lambda\) be a linear functional on the Cartan subalgebra \(\mathfrak{h}\). The \textit{weight space} of \(V\) with weight \(\lambda\) is
\begin{equation}
    V_{\lambda} = \qty{v \in V \mid \forall H \in \mathfrak{h}, \; H \cdot v = \lambda(H) v}.
\end{equation}
Essentially, this generalizes the familiar ladder operators of \(\su(2)\). As a consequence of the root equation \eqref{eq:rootequation} if \(v\) is a weight vector with weight \(\lambda\)
\begin{equation}
\begin{split}
    H \cdot X v =& \; X \cdot H v + \qty[H, X] v\\[2mm]
    =& \; \qty[\lambda(H) + \alpha(H)] X v, \quad \forall H \in \mathfrak{h}.
\end{split}
\end{equation}
Therefore \(X v\) is either the zero vector or a weight vector with weight \(\lambda + \alpha\).

Denote by \(\mathfrak{h}^{*}_{0}\) the real subspace of the dual of the Cartan subalgebra \(\mathfrak{h}^{*}\) generated by the roots of \(\mathfrak{g}\). Then an element \(\lambda \in \mathfrak{h}_{0}\) is said to be \textit{algebraically integral} if 
\begin{equation}
    \expval{\lambda, \alpha^{\vee}_{jk}} \in \mathbb{Z}, \quad \forall \alpha^{\vee}_{jk} \in \mathfrak{h}^{*}_{0}.
\end{equation}
The coroot \(\alpha^{\vee}_{jk}\) can be identified with the \(H\) element in the \(X\), \(Y\), \(H\) basis of an \(\spl(2, \mathbb{C})\)-subalgebra of \(\mathfrak{g}\). The eigenvalues of \(\alpha^{\vee}_{jk}\) in any finite dimensional representation of \(\mathfrak{g}\) must then be an integer.

For an algebraically integral element \(\lambda\) we can write its components in the \(\omega\)-basis
\begin{equation}
    a_{j} = \expval{\lambda, \alpha^{\vee}_{j}},
\end{equation}
where the \(a_{j}\) are integers for every simple root. The smallest non-zero weights with \(a_{j} \geq 0\) are the fundamental weights. We then refer to the \(a_{j}\) as the \textit{Dynkin-labels} of the weight. Therefore, an element \(\lambda\) is algebraically integral if and only if it is expressible as a combination of the fundamental weights with integer coefficients. A weight is called \textit{dominant} if all of the \(a_{j} \geq 0\).

The weights of a representation can be partially ordered. Consider the positive roots in \(\Delta\), denoted \(\Delta^{+}\), as the set of all \(\alpha^{j k}\) with \(j < k\). We then partially order the weights as
\begin{equation}
    \lambda \succeq \mu \qq{if} \lambda - \mu = \theta_{i j} \alpha^{i j}, \quad \theta_{i j} \in \mathbb{R}^{+}, \alpha^{i j} \in \Delta^{+}.
\end{equation}
That is, \(\lambda\) is higher than \(\mu\) if the difference \(\lambda - \mu\) can be written as a linear combination of positive roots with non-negative real coefficients. A weight \(\lambda\) of a representation \(V\) of an algebra \(\mathfrak{g}\) is then said to be the \textit{highest weight} if every other weight of \(V\) is lower than \(\lambda\). The \textit{level} of a weight is the number of simple roots that must be subtracted from the highest weight in order to obtain it. The highest level of an \irrep{} is call its \textit{height}. We can now state the following theorem:
\begin{theorem}{Theorem of the Highest Weight} \label{thm:highestweight}
\begin{enumerate}
    \item Every irreducible (finite dimensional) representation has a highest weight,
    \item the highest weight is always a dominant, algebraically integral element,
    \item two \irreps{} with the same highest weight are isomorphic, and
    \item every dominant, algebraically integral element is the highest weight of an \irrep{}.
\end{enumerate}
\end{theorem}

\begin{figure}[ht!]
\centering
\includegraphics[width=\columnwidth]{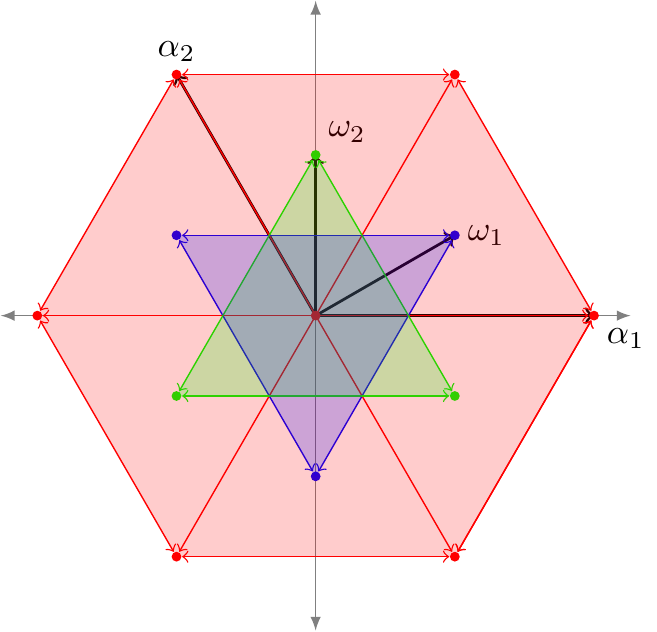}
\caption{Weight spaces for some \irreps{} of \(\SU(3)\). Shown are the fundamental (\(3\)), antifundamental (\(\bar{3}\)), and adjoint (\(8\)) representations in blue, green, and red respectively. \label{fig:weightspaces}}
\end{figure}

\begin{figure}[ht!]
\centering
\includegraphics[width=\columnwidth]{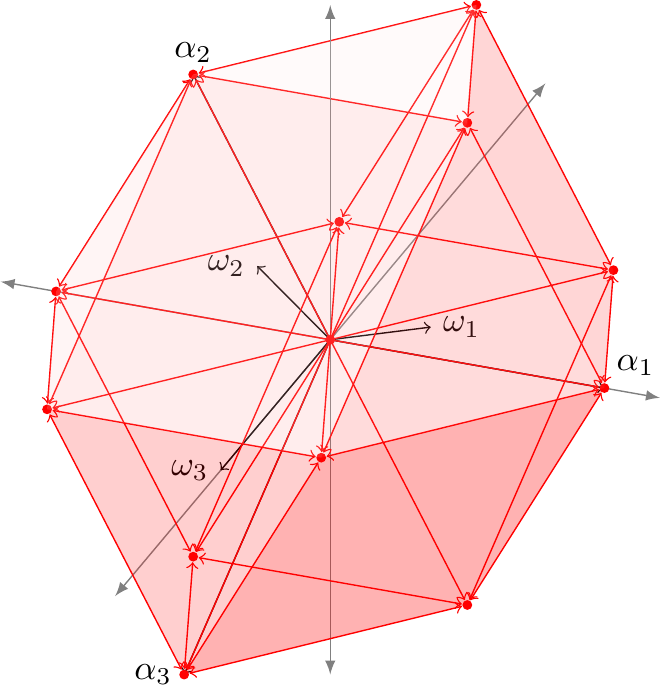}
\caption{Weights space of the adjoint representation in \(\SU(4)\). Vertices indicate locations of the weights while red arrows indicate the Weyl group orbits.\label{fig:weightspacesu4}}
\end{figure}

Finally, a finite-dimensional representation \(V\) of a semisimple Lie algebra \(\mathfrak{g}\) is uniquely specified by a set of weights. If \(V\) is generated by a \(v \in V\) that is annihilated by the action of all positive root spaces in \(\mathfrak{g}\) then \(V\) is called a \textit{highest-weight module}. It can be shown that every finite-dimensional highest-weight module is irreducible. A representation can then be decomposed into \irreps{} by sorting the weights into weight spaces that correspond to the \irreps{}. The dimension of a representation is just the number of weights in its weight space. Several examples of weights spaces are given in Fig.~\ref{fig:weightspaces} and Fig.~\ref{fig:weightspacesu4}.

\section{Proof of Adjoint Product Decomposition} \label{app:adjointprod}

The vector space of the adjoint representation of \(\SU(N)\) is spanned by the generators of the algebra of \(\su(n)\). There are \(N-1\) zero weights coming from the \(N-1\) generators of the Cartan subalgebra, and \(N \qty(N-1)\) weights from the remaining generators with roots \(\alpha^{jk}\), giving the dimension of the adjoint representation as \(N^{2} - 1\) since there are \(N^{2} - 1\) weights. By Theorem~\ref{thm:highestweight} there is a unique highest weight,
\begin{equation}
    \bar{\lambda} = \alpha^{1 N} = \sum^{N-1}_{j = 1} \alpha^{j} = \omega^{1} + \omega^{N-1},
\end{equation}
which we use to label the adjoint representation. Taking the tensor product of \(2\) copies of the adjoint we add all the weights of one copy of the adjoint to every weight of the second copy. This gives a set of \(\qty(N^{2} - 1) \qty(N^{2} - 1) \) weights, many of them duplicates, that must be sorted in order to decompose the tensor product into a direct sum of \irreps{}. Continuing to take the tensor product \(k\) times in total produces \(\qty(N^{2} - 1)^{k}\) weights, and sorting all of these weights quickly becomes unreasonably tedious. For fixed \(k\) it can be done using Freudenthal's recursion formula for the multiplicities of the weight \(\lambda\) in the \irrep{} with highest weight \(\Lambda\):
\begin{multline}
    \qty(\expval{\Lambda + \delta, \Lambda + \delta} - \expval{\lambda + \delta, \lambda + \delta}) m_{\Lambda}(\lambda)\\[2mm]
    = 2 \sum_{\alpha \in \Delta^{+}} \sum_{j \geq 1} \expval{\lambda + j \alpha, \alpha} m_{\Lambda}(\lambda + j \alpha),
\end{multline}
where \(\expval{\cdot, \cdot}\) is the usual inner product on \(\mathbb{R}^{N}\), \(\delta \coloneqq \sum^{N-1}_{j = 1} \omega^{j}\) is the Weyl vector, \(m_{\Lambda}(\lambda)\) is the multiplicity of the weight \(\lambda\) in the \irrep{} with highest weight \(\Lambda\), and the first sum is over all positive roots of the algebra. If \(k\) is left as a free parameter this formula becomes less useful, so we take a different approach based on Schur-Weyl duality.

The tensor product of any two \irreps{} of \(\SU(N)\), \(V_{\mu}\) and \(V_{\nu}\), can be accomplished in the following way:
\begin{equation}
    V_{\mu} \otimes V_{\nu} \cong \bigoplus_{\lambda} c^{\lambda}_{\mu \nu} V_{\lambda},
\end{equation}
where \(c^{\lambda}_{\mu \nu}\) is the multiplicity of the \irrep{} labeled by its highest weight \(\lambda\) in the tensor product and the sum is taken over all dominant weights \(\lambda\). The multiplicities are given by the \textit{Littlewood-Richardson rule}, which counts the number of Littlewood-Richardson tableau with skew shape \(\lambda / \mu\) and content \(\nu\). They appear as the coefficients in the decomposition of the multiplication of Schur functions
\begin{equation}
    s_{\mu} s_{\nu} = \sum_{\lambda} c^{\lambda}_{\mu \nu} s_{\lambda}.
\end{equation}
This decomposition follows from the Weyl character formula for representations of a simple Lie algebra over a complex field\cite{steinberg1961}. In fact, the multiplicities can be calculated explicitly, without reference to Littlewood-Richardson tableau:
\begin{theorem}{Steinberg \cite{steinberg1961}} \label{thm:steinberg}
\begin{equation}
c^{\lambda}_{\mu \nu} = \sum_{\sigma, \tau \in \mathfrak{S}_{N}} \det(\sigma \tau) \mathfrak{p}(\sigma(\mu + \delta) + \tau(\nu + \delta) - \qty(\lambda + 2 \delta)), 
\end{equation}
where \(\mathfrak{S}_{N}\) is the symmetric group of permutations acting on vectors with \(N\) elements, \(\delta\) is the Weyl vector, and \(\mathfrak{p}(v)\) is the Kostant partition function.
\begin{equation}
    \mathfrak{p}(v) = \abs{\qty{n_{\alpha} \in \mathbb{R}^{N-1} \mid v = \sum_{\alpha \in \Delta^{+}} n_{\alpha} \alpha, n_{\alpha} \geq 0}}.
\end{equation}
\(\mathfrak{p}(v)\) counts the number of ways a vector \(v\) (weight) can be written as the sum of positive roots.
\end{theorem}
In particular, if we let either \(\mu\) or \(\nu\) be either \(\qty(1, 0, \ldots, 0)\) or \(\qty(1,1,\ldots,1,0,\ldots,0)\) we have Pieri's formula:
\begin{equation}
    s_{\mu} h_{r} = \sum_{\lambda} s_{\lambda},
\end{equation}
where \(h_{r}\) is a complete homogeneous symmetric polynomial and the sum is taken over all partitions \(\lambda\) obtained from \(\mu\) by adding \(r\) boxes with no boxes in the same column. We also have the dual Pieri rule
\begin{equation}
    s_{\mu} e_{r} = \sum_{\lambda} s_{\lambda},
\end{equation}
where \(e_{r}\) is an elementary symmetric polynomial and the sum is taken over all partitions \(\lambda\) obtained from \(\mu\) by adding \(r\) boxes with no boxes in the same row. For these choices of the partitions \(\mu\) and \(\nu\) the Littlewood-Richardson coefficients are either 1 or 0.

Schur-Weyl duality comes from considering the \(k\)-fold tensor product of \(\mathbb{C}^{N}\). The action of the symmetric group \(\mathfrak{S}_{k}\) and the general linear group \(G = \GL(\mathbb{C}^{N})\) commute, and as an \(\mathfrak{S}_{k} \times G\)-module we have the decomposition
\begin{equation}
\qty(\mathbb{C}^{N})^{\otimes k} \cong \bigoplus_{\lambda} M_{\lambda} \otimes \mathbb{S}_{\lambda} \mathbb{C}^{N},
\end{equation}
where \(M_{\lambda}\) is an \irrep{} of \(\mathfrak{S}_{N}\) indexed by the partition \(\lambda\) and the Schur functor \(\mathbb{S}_{\lambda} V\) is the image of the Young symmetrizer. The fundamental representation, \(V_{F}\), of \(\SU(N)\) is a \(G\)-module isomorphic to \(\mathbb{C}^{N}\), hence we have
\begin{equation}
    V^{\otimes k}_{F} \cong \bigoplus_{\lambda} V_{\lambda}^{\oplus m_{\lambda}},
\end{equation}
where \(m_{\lambda} \equiv \dim(M_{\lambda})\), and \(V_{\lambda} \equiv \mathbb{S}_{\lambda} \mathbb{C}^{N}\) is an \irrep{} of \(\SU(N)\). The sum is over all partitions \(\lambda\) of the integer \(k\). In addition, the dual representation to \(V_{F}\), denoted \(V^{*}_{F}\), is also isomorphic to \(\mathbb{C}^{N}\), so we also have
\begin{equation}
    \qty(V^{*}_{F})^{\otimes k} \cong \bigoplus_{\lambda} V_{\lambda^{*}}^{\oplus m_{\lambda}},
\end{equation}
where, again, the sum is over all partitions \(\lambda\) of the integer \(k\) and \(\lambda^{*}\) is the conjugate partition to \(\lambda\).

Using Pieri's rule we can relate the adjoint representation to the fundamental and antifundamental representations of \(\SU(N)\):
\begin{equation}
    V_{F} \otimes V^{*}_{F} \cong G_{N} \oplus \mathbbm{1}.
\end{equation}
This then implies
\begin{equation} \label{eq:FundAntiFundprod_k}
    \qty(V_{F} \otimes V^{*}_{F})^{\otimes k} \cong \bigoplus^{k}_{n = 0} \binom{k}{n} G^{\otimes k - n}_{N},
\end{equation}
where \(G^{\otimes 0}_{N} \equiv \mathbbm{1}\). Therefore, we have the implications
\begin{equation} \label{eq:implication1}
    V_{\lambda} \subset G^{\otimes n}_{N} \implies V_{\lambda} \subset \qty(V_{F} \otimes V^{*}_{F})^{\otimes k}, \quad \forall n \leq k,
\end{equation}
and
\begin{equation} \label{eq:implication2}
    V_{\lambda} \subset \qty(V_{F} \otimes V^{*}_{F})^{\otimes k} \implies \exists\, n \mid 0 \leq n \leq k, \; V_{\lambda} \subset G^{\otimes n}_{N}. 
\end{equation}
We then look at the tensor product of two copies of the adjoint:
\begin{equation}
    G_{N} \otimes G_{N} \cong \bigoplus_{\lambda} c^{\lambda}_{\bar{\lambda} \bar{\lambda}} V_{\lambda},
\end{equation}
where \(\bar{\lambda} = \qty(1, 0, \ldots, 0, 1)\) is the highest weight of \(G_{N}\). Using Theorem~\ref{thm:steinberg} we explicitly calculate the multiplicities
\begin{subequations}
\begin{align}
    c^{\bar{\lambda}}_{\bar{\lambda} \bar{\lambda}} =& \; \begin{cases} 1 & N = 2\\ 2 & N > 2 \end{cases},\\[2mm]
    c^{0}_{\bar{\lambda} \bar{\lambda}} =& 1.
\end{align}
\end{subequations}
This shows that \(G_{N} \subset G_{N} \otimes G_{N}\) and \(\mathbbm{1} \subset G_{N} \otimes G_{N}\). It is then always the case that
\begin{equation}
    G^{\otimes n - 1}_{N} \subset G^{\otimes n}_{N}, \quad \forall n > 1.
\end{equation}
But then from \eqref{eq:implication2} we find
\begin{equation} \label{eq:inclusion}
    V_{\lambda} \subset \qty(V_{F} \otimes V^{*}_{F})^{\otimes k} \implies V_{\lambda} \subset G^{\otimes k}_{N}, \quad \forall k > 1.
\end{equation}
We can then invert \eqref{eq:FundAntiFundprod_k} to solve for \(G^{\otimes k}_{N}\):
\begin{equation} \label{eq:GkN}
\begin{split}
    G^{\otimes k}_{N} \cong& \; \qty(V_{F} \otimes V^{*}_{F} \ominus \mathbbm{1})^{\otimes k}\\[2mm]
    \cong& \; \bigoplus^{k}_{n = 0} \qty(-1)^{k - n} \binom{k}{n} V^{\otimes n}_{F} \otimes \qty(V^{*}_{F})^{\otimes n}\\[2mm]
    \cong& \; \bigoplus^{k}_{n = 0} \qty(-1)^{k - n} \binom{k}{n} \bigoplus_{\mu_{n}} m_{\mu_{n}} V_{\mu_{n}} \bigoplus_{\nu_{n}} m_{\nu_{n}} V_{\nu^{*}}\\[2mm]
    \cong& \bigoplus_{\lambda} \bigoplus^{k}_{n = 0} \bigoplus_{\mu_{n}} \bigoplus_{\nu_{n}} \qty(-1)^{k - n} \binom{k}{n} m_{\mu_{n}} m_{\nu_{n}} c^{\lambda}_{\mu_{n} \nu^{*}_{n}} V_{\lambda}\\[2mm]
    \cong& \bigoplus_{\lambda} V_{\lambda}^{\oplus m_{\lambda}(G_{N})}.
\end{split}
\end{equation}
The sum in the last line of \eqref{eq:GkN} is over all dominant \(\lambda\), and the multiplicity of the \irrep{} \(\lambda\) in the decomposition is given by
\begin{equation}
    m_{\lambda}(G_{N}) = \sum^{k}_{n = 0} \sum_{\mu_{n}} \sum_{\nu_{n}} \qty(-1)^{k - n} \binom{k}{n} m_{\mu_{n}} m_{\nu_{n}} c^{\lambda}_{\mu_{n} \nu^{*}_{n}},
\end{equation}
where the second and third sums are over all partitions \(\mu_{n}\) and \(\nu_{n}\) of the integer \(n\) and \(\nu^{*}_{n}\) is the conjugate partition to \(\nu_{n}\). The highest weight of the highest \irrep{} in \(G^{\otimes k}_{N}\) is
\begin{equation}
    \Lambda \equiv k \bar{\lambda} = k \alpha^{1N} = k \omega^{1} + k \omega^{N-1}.
\end{equation}
As a consequence of \eqref{eq:inclusion} we have the following identity:
\begin{equation} \label{eq:m_lam_GN}
    m_{\lambda}(G_{N}) > 0, \quad \forall \lambda \text{ dominant} \mid \lambda \preceq \Lambda.
\end{equation}

Because of \eqref{eq:m_lam_GN} every dominant weight \(\lambda\) will show up as an \irrep{} in the tensor product \(G^{\otimes k}_{N}\). As previously stated the multiplicity is irrelevant for our purposes, we only need \(m_{\lambda}(G_{N}) > 0\). Since every dominant \(\lambda\) in the tensor product can be written as
\begin{equation}
    \lambda = \Lambda - \sum^{N-1}_{j = 1} n_{j} \alpha^{j},
\end{equation}
and since every positive root can be written as a sum of simple roots
\begin{equation}
    \lambda = \Lambda - \sum^{N}_{j = 1} \sum_{k > j} i_{j k} \alpha^{j k}, \quad \forall V_{\lambda} \subset G^{\otimes k}_{N}.
\end{equation}
Then writing the components of this weight in the \(\omega\) basis as
\begin{equation}
    p_{j} \coloneqq \expval{\lambda, \alpha^{\vee}_{j}},
\end{equation}
we see that this gives precisely the conditions given in Section~\ref{subsec:AdProd} with the definition
\begin{equation}
    \Delta p_{j} \coloneqq \sum^{N}_{k = 1} \sum_{l > k} i_{j k} \expval{\alpha^{kl}, \alpha^{\vee}_{j}}.
\end{equation}
Indeed, it is easy to show that, because
\begin{equation}
\begin{split}
    \expval{\alpha^{k l}, \alpha^{\vee}_{j}} =& \; \sum^{N}_{n = 1} \qty(\delta^{k}_{n} - \delta^{l}_{n}) \qty(\delta^{n}_{j} - \delta^{n}_{j + 1}),\\[2mm]
    =& \; \delta^{k}_{j} - \delta^{l}_{j} - \qty(\delta^{k}_{j+1} - \delta^{l}_{j + 1}),
\end{split}
\end{equation}
we have exactly our definitions of the integers \(m_{j}\) as before:
\begin{equation}
    m_{j} = \sum^{N}_{k = 1} \underbrace{\sum_{l > k} i_{k l} \qty(\delta^{k}_{j} - \delta^{l}_{j})}_{M_{j k}}.
\end{equation}
Because every \(\lambda\) of this form has non zero multiplicity our procedure in Section~\ref{subsec:AdProd} is complete.

\section{User Manual for the {\tt tessellation} code}\label{app:code}

The code can be easily installed by cloning the repository \url{https://github.com/jaulbric/Tesselation}. Alternatively, the package can be downloaded by issuing the following commands via a terminal:

\vspace{0.5cm}

\noindent\underline{\textbf{bash}}:
\begin{minted}[breaklines, breakafter=/]{bash}
# Get the version of the latest release
version=`curl -s https://api.github.com/repos/jaulbric/Tesselation/releases/latest \
| grep "tag_name" \
| cut -d ":" -f 2 \
| tr -d " "-\"-,`
# download and untar the package
wget https://github.com/jaulbric/Tesselation/archive/$version.tar.gz \
&& mkdir ~/Tesselation \
&& tar xvfz $version.tar.gz -C ~/Tesselation \
--strip-components 1 && rm $version.tar.gz
\end{minted}

\noindent\underline{\textbf{csh/tcsh}}:
\begin{minted}[breaklines, breakafter=/]{csh}
# Get the version of the latest release
set version=`curl -s https://api.github.com/repos/jaulbric/Tesselation/releases/latest \
| grep "tag_name" \
| cut -d ":" -f 2 \
| tr -d " "-\"-,`
# download and untar the package
wget https://github.com/jaulbric/Tesselation/archive/$version.tar.gz \
&& mkdir ~/Tesselation \
&& tar xvfz $version.tar.gz -C ~/Tesselation \
--strip-components 1 && rm $version.tar.gz
\end{minted}
Once downloaded the package can be installed with \mintinline{bash}{pip install .} in the root directory of the repository (in the above examples \url{~/Tesselation}).

We now give an example of how to use the code by reproducing the results of Fig.~\ref{fig:matterfieldexamples}.
\begin{minted}{python}
import numpy as np
import tesselation as t

X1 = np.array([3, 0, 0, 1])
X2 = np.array([0, 2, 1, 0])
X3 = np.array([0, 0, 1, 1])
X4 = np.array([1, 3, 0, 0])

N = 5
i = N - t.Nality(X1) # This is 3 for all cases

idx = 1
for x in [X1, X2, X3, X4]:
    l1, l2 = t.l_pair(x, i) # Calculate l1 and l2
    fj = t.fj(x) # Calculate the Dynkin label
    kmin = t.kmin(x) # Calculate kmin
    print("X{0}:".format(idx))
    print("l1 = {0}, l2 = {1}".format(l1, l2))
    print("f = ({0}, {1}, {2}, {3})".format(*fj))
    print("kmin = {0}\n".format(kmin))
    idx += 1
\end{minted}
The output of the above code is
\begin{minted}{text}
X1:
l1 = 1, l2 = 4
f = (2, 0, 1, 0)
kmin = 2

X2:
l1 = 2, l2 = 3
f = (0, 1, 1, 0)
kmin = 2

X3:
l1 = 0, l2 = 3
f = (1, 0, 0, 1)
kmin = 1

X4:
l1 = 2, l2 = 5
f = (1, 2, 0, 0)
kmin = 3
\end{minted}

\noindent There are four helper functions available to the user:
\begin{itemize}
    \item \mintinline{python}{l_pair(p, i)}: Takes a Dynkin label (\mintinline{python}{list} or \mintinline{python}{numpy.ndarray}) and a \mintinline{python}{int} as input and outputs \(\ell_{1}\) and \(\ell_{2}\) as described in Section~\ref{subsec:AdProd}.
    \item \mintinline{python}{Nality(p)}: Takes a Dynkin label (\mintinline{python}{list} or \mintinline{python}{numpy.ndarray}) as an input and returns the $N$-ality of the representation, that is \(\sum^{N-1}_{j = 0} j p_{j} \mod{N}\).
    \item \mintinline{python}{fj(p)}: Takes a Dynkin label (\mintinline{python}{list} or \mintinline{python}{numpy.ndarray}) as an input and returns the Dynkin label of the \irrep{} in \(X \otimes Q_{i}\) that gives the smallest value of \(k_{\mathrm{min}}\).
    \item \mintinline{python}{kmin(p)}: Takes a Dynkin label (\mintinline{python}{list} or \mintinline{python}{numpy.ndarray}) as an input and returns the minimum number of copies of the adjoint representation required such that \(\mathbbm{1} \subset X \otimes Q_{i} \otimes G^{\otimes k_{\mathrm{min}}}_{N}\).
\end{itemize}

\nocite{*}
\bibliography{bibliography}% Produces the bibliography via BibTeX.

\end{document}